\documentclass[12pt]{article}
\pdfoutput=1
\usepackage{fullpage}
\usepackage{color}
\usepackage{latexsym}
\usepackage{mathrsfs}
\usepackage{amsfonts}
\usepackage{booktabs}
\usepackage{amsmath,amssymb}
\usepackage{bbm} %for \mathbbm{1} and \mathbbm{b}
\usepackage[toc]{appendix}
\usepackage[numbers,compress]{natbib}

\usepackage[T1]{fontenc}

\usepackage{amssymb,amsmath,amscd}
\usepackage{epsfig}
\usepackage{epstopdf}
\usepackage{latexsym}
\usepackage{graphicx}
\usepackage{subfigure}
\usepackage{booktabs}
\usepackage{bbm}

\usepackage{color}
\usepackage{datetime}
\usepackage[
      colorlinks=false,
      linkcolor=darkblue,
      urlcolor=blue,
      filecolor=blue,
      citecolor=red,
linktocpage=true,
      pdfstartview=FitV,
      bookmarksopen=true
      ]{hyperref}

\numberwithin{equation}{section}       % equation numbers in each section
\newcommand{\be}{\begin{equation}}
\newcommand{\ee}{\end{equation}}
\newcommand{\bea}{\begin{eqnarray}}
\newcommand{\eea}{\end{eqnarray}}
\newcommand{\nn}{\nonumber}

\newcommand\e{\mathrm{e}}
\newcommand\iu{\operatorname{i}}
\def\vct{X}

\newcommand{\diff}{\mathrm{d}}

\newcommand{\rme}{\mathrm{e}}

\newcommand{\ch}{\rm ch}
\newcommand{\sh}{\rm sh}

\newcommand{\SO}{{\rm SO}}
\newcommand{\SU}{{\rm SU}}

\newcommand{\U}{{\rm U}}
\newcommand{\USp}{{\rm USp}}

\newcommand{\SL}{{\rm SL}}

\def\cV{\mathcal{V}}

\def\SO{\textrm{SO}}
\def\SU{\textrm{SU}}

\DeclareRobustCommand{\SkipTocEntry}[4]{}
\RequirePackage{color}
\definecolor{cardinal}{rgb}{0.6,0,0}
\definecolor{darkgreen}{rgb}{0,0.5,0}
\definecolor{golden}{rgb}{0.92, 0.7, 0}
\definecolor{midnight}{rgb}{0, 0, 0.5}
\definecolor{darkblue}{rgb}{0.2, 0, 0.8}

\begin{document}

%%%%%%%%%%%%%%%%%%%%%%%%%%%%%%%%%%%%%%%%%%
\begin{titlepage}
%%%%%%%%%%%%%%%%%%%%%%%%%%%%%%%%%%%%%%%%%%

\begin{center}

\rightline{\small Imperial/TP/18/HT/01}

\vskip 2cm

{\Large \bf Holographic RG Flows for Four-dimensional $\mathcal{N}=2$~SCFTs}

\vskip 1.7cm

{\bf Nikolay Bobev$^a$, Davide Cassani$^b$ and Hagen Triendl$^c$} \\

\vskip 1.2cm

{}$^{a}${\em Instituut voor Theoretische Fysica, KU Leuven,\\
Celestijnenlaan 200D, B-3001 Leuven, Belgium
\vskip 5mm}

{}$^{b}${\em  INFN, Sezione di Padova, Via Marzolo 8, 35131 Padova, Italy
\vskip 5mm}

{}$^{c}${\em Department of Physics, Imperial College London,\\
Prince Consort Road, London SW7 2AZ, UK}

\vskip 0.8cm

{\tt nikolay.bobev@kuleuven.be, davide.cassani@pd.infn.it, h.triendl@imperial.ac.uk} \\

\end{center}

\vskip 1cm

\begin{center} {\bf Abstract } \end{center}

\noindent We study holographic renormalization group flows from four-dimensional $\mathcal{N}=2$ SCFTs to either $\mathcal{N}=2$ or $\mathcal{N}=1$ SCFTs. Our approach is based on the framework of five-dimensional half-maximal supergravity with general gauging, which we use to study domain wall solutions interpolating between different supersymmetric AdS$_5$ vacua. We show that a holographic RG flow connecting two $\mathcal{N}=2$ SCFTs is only possible if the flavor symmetry of the UV theory admits an $\SO(3)$ subgroup. In this case the ratio of the IR and UV central charges satisfies a universal relation which we also establish in field theory.
In addition we provide several general examples of holographic flows from $\mathcal{N}=2$ to $\mathcal{N}=1$ SCFTs and relate the ratio of the UV and IR central charges to the conformal dimension of the operator triggering the flow. Instrumental to our analysis is a derivation of the general conditions for AdS vacua preserving eight supercharges as well as for domain wall solutions preserving eight Poincar\'e supercharges in half-maximal supergravity.

\vfill

%%%%%%%%%%%%%%%%%%%%%%%%%%%%%%%%%%%%%%%%%
\end{titlepage}
%%%%%%%%%%%%%%%%%%%%%%%%%%%%%%%%%%%%%%%%%

\tableofcontents

%%%%%%%%%%%%%%%%%%%%%%%%%%%%%%%%%%%%%%%%%%%%%%%%%%%%%%%%%%%%%%%%%%%%%%%%%%%%%%%%%%%%%%
\section{Introduction}

The AdS/CFT correspondence has offered many insights into the structure of the Renormalization Group (RG). Soon after the original conjecture by Maldacena it was realized that one can use supersymmetric gravitational domain wall solutions interpolating between AdS$_5$ vacua of string theory to describe  RG flows between strongly interacting four-dimensional superconformal field theories (SCFTs) \cite{Girardello:1998pd,Freedman:1999gp}. Furthermore this vantage point was used to obtain a holographic proof of the $a$-theorem. There have been numerous generalizations of these original results to SCFTs in other dimensions. Our goal here is to build upon this body of work and study general constraints on supersymmetric holographic RG flows.  Rather than studying explicit top-down models in ten- or eleven-dimensional string/M-theory, we will explore the space of AdS vacua and the flows connecting them in general gauged supergravity. In this paper our main interest is in RG flows deforming four-dimensional $\mathcal{N}=2$ SCFTs, hence we will work in half-maximal supergravity in five dimensions (though we will also extend some of our results to other dimensions).

This ``bottom-up'' approach to studying general properties of SCFTs with holographic duals by employing gauged supergravity has recently received some attention in the context of understanding the space of exactly marginal couplings of SCFTs in various dimensions, see \cite{Hristov:2009uj,Louis:2009xd,deAlwis:2013jaa,Louis:2014gxa,Louis:2015mka,Louis:2015dca,Louis:2016qca,Karndumri:2016ruc,Lust:2017aqj,Lust:2017fyw} and references thereof. While these results present interesting holographic predictions for strongly interacting SCFTs it should be noted that some of the supergravity results can also be shown more generally, without a reference to holography, using purely algebraic techniques \cite{Cordova:2016xhm}. Questions about the structure of RG flows triggered by relevant deformations on the other hand are less accessible with field theory tools and thus the supergravity results  derived here should teach us important general lessons for the structure of supersymmetric RG flows.

To understand the general constraints for the existence of distinct supersymmetric AdS$_5$ vacua and the flow connecting them we present a detailed analysis of the supersymmetry conditions in half-maximal gauged supergravity. The results depend on the number, $\mathfrak{n}$, of vector multiplets in the theory and on the type of gauging performed. The existence of at least one AdS$_5$ vacuum with 16 supercharges implies that an $\U(1)\times \SU(2) \times H_{\rm c}$ subgroup of the $\SO(5,\mathfrak{n})$ global symmetry of the supergravity theory should be gauged \cite{Louis:2015dca}. The $\U(1)\times \SU(2)$ gauge field is dual to the R-symmetry of the four-dimensional $\mathcal{N}=2$ SCFT dual to this AdS$_5$ while $H_{\rm c}$ represents the continuous flavor symmetry. If $H_{\rm c}$ is trivial we find that there is a unique AdS$_5$ vacuum with 16 supercharges in the supergravity theory.\footnote{This result can also be established for AdS vacua with 16 supercharges in four-, six-, and seven-dimensional half-maximal gauged supergravity.} However when $H_{\rm c}$ is non-trivial then it must contain an $\SO(3)$ subgroup and there can be another AdS$_5$ vacuum in the supergravity theory with a different value of the cosmological constant. Moreover these two distinct AdS$_5$ vacua are connected by a regular supersymmetric domain wall solution in the gauged supergravity theory which we construct analytically. In addition we establish that the RG flow in the dual QFT should be triggered by vacuum expectation values (vevs) for two scalar operators of dimension $\Delta=2$ and the ratio of these vevs has to be a fixed constant. One of the two scalar operators belongs to the energy momentum multiplet of the SCFT and the other one sits in the $\SO(3)\subseteq H_{\rm c}$ flavor current multiplet. The different values of the cosmological constants of the two AdS$_5$ vacua translate into different values for the conformal anomalies of the dual UV and IR $\mathcal{N}=2$ SCFTs. We compute this ratio of central charges using our supergravity results and are able to reproduce it by an anomaly calculation in the dual SCFT. The result is a universal expression for the IR conformal anomalies in terms of the UV conformal anomalies as well as the central charges of the $\SO(3)$ flavor current. We also find that these anomalies are related to the constant that controls the relation between the scalar vevs triggering the flow.

Having described the conditions for the existence of $\mathcal{N}=4$ AdS vacua in five-dimensional gauged supergravity it is natural to ask whether there are other AdS vacua which preserve less supersymmetry. To answer this we analyze the general conditions for $\mathcal{N}=2$ AdS$_5$ vacua and then we focus on theories that admit both an $\mathcal{N}=4$ and one or more $\mathcal{N}=2$ vacua. Perhaps not surprisingly we find that as we increase the number of vector multiplets in the supergravity theory we can have an increasing number of distinct $\mathcal{N}=2$ AdS$_5$ vacua. The details of this structure depend on the matter content and the choice of gauging in the supergravity theory. To illustrate our general approach we focus on two particular examples. We first establish a holographic analog of the QFT result in \cite{Tachikawa:2009tt} in which it was shown that every four-dimensional $\mathcal{N}=2$ SCFT with an exactly marginal deformation admits an RG flow to an IR $\mathcal{N}=1$ SCFT. In addition it was shown in \cite{Tachikawa:2009tt} that the conformal anomalies of the IR and UV SCFTs obey a universal relation. This type of RG flow was in fact first constructed and discussed in some particular holographic examples, see \cite{Freedman:1999gp,Corrado:2002wx,Corrado:2004bz}, but here we offer a more general treatment. Our general setup should capture the RG flow relating the $\mathcal{N}=2$ and $\mathcal{N}=1$ Maldacena-N\'u\~nez SCFTs \cite{Maldacena:2000mw} arising from M5-branes wrapped on a Riemann surface. While it is widely believed that this RG flow exists, and is of the class discussed in \cite{Tachikawa:2009tt}, its explicit holographic construction is still elusive. Our results should offer some insight into this problem. Moreover, if our setup can be embedded in eleven-dimensional supergravity it can potentially capture holographic RG flows connecting the $\mathcal{N}=2$ Maldacena-N\'u\~nez SCFT \cite{Maldacena:2000mw} and one of the $\mathcal{N}=1$ SCFTs studied in \cite{Bah:2011vv,Bah:2012dg}. In addition to this we study a setup with one $\mathcal{N}=4$ and two distinct $\mathcal{N}=2$ AdS$_5$ vacua and discuss the supersymmetric domain wall solutions which interpolate between them. This may capture holographic RG flows which relate the $\mathcal{N}=2$ Maldacena-N\'u\~nez SCFT and two of the $\mathcal{N}=1$ SCFTs of \cite{Bah:2011vv,Bah:2012dg}.

Finally we would like to note that we do not study a specific embedding of the gauged supergravity theories we work with in string or M-theory. Thus our results are universal and apply to all supersymmetric AdS vacua which admit a lower-dimensional effective description in terms of half-maximal supergravity. This universality is somewhat similar in spirit to the results for holographic RG flows across dimensions discussed in \cite{Bobev:2017uzs}.

We begin our presentation in the next section with a brief general introduction to five-dimensional $\mathcal{N}=4$ gauged supergravity. In Section~\ref{sec:onlyN=4vacua} we identify under what conditions there can be two distinct AdS vacua of such a supergravity theory which preserve all 16 supercharges and construct gravitational domain wall solutions interpolating between these vacua.  Whenever such a flow is possible it exhibits a universal relation between the UV and IR central charges which we establish by field theory methods in Section~\ref{sec:QFT}. We continue in Section~\ref{sec:Flow} with a study of the conditions for the existence of AdS$_5$ vacua with 8 supercharges and a discussion on domain wall solutions connecting such vacua. Section~\ref{sec:Discussion} is devoted to a short discussion on our results and their implications for holography. In Appendix~\ref{app:uniqueness4d} we present the extension of some of the results in Section~\ref{sec:onlyN=4vacua} to half-maximal gauged supergravity in four, six and seven dimensions. In Appendix~\ref{app:genTW} we give some more details on the flow in Section~\ref{sec:Flow}.

%%%%%%%%%
\section{Gauged half-maximal supergravity}
\label{sec:Prelim}
%%%%%%%%%

In this section we review the basic properties of five-dimensional gauged $\mathcal{N}=4$ (half-maximal) supergravity \cite{Romans:1985ps,Awada:1985ep,Dall'Agata:2001vb,Schon:2006kz} that are relevant for our analysis, mainly following~\cite{Schon:2006kz}.

Ungauged $\mathcal{N}=4$ supergravity  has $\USp(4)$ R-symmetry and consists of a gravity multiplet and $\mathfrak{n}$~vector
multiplets.  The gravity multiplet contains the metric $g_{\mu\nu}$, four
gravitini $\psi^i_{\mu},\, i=1,\ldots,4$ transforming in the ${\bf 4}$ of $\USp(4)$, six vectors (dubbed the graviphotons) $A_\mu^0, A_\mu^m$, with $A_\mu^m$, $m=1,\ldots, 5$ transforming in the ${\bf 5}$ of $\USp(4)$ and $A_\mu^0$ being neutral,
four spin-1/2 fermions $\chi^i$ in the ${\bf 4}$ of $\USp(4)$, and one neutral real scalar $\Sigma$. We will label the vector multiplets with the index $a=1,\ldots,\mathfrak{n}$. Each vector multiplet contains a vector $A_{\mu}^a$, four spin-1/2 gaugini $\lambda^{ai}$, and five real scalar fields. All together the scalar fields parametrize the coset space
\be\label{N4coset}
\mathcal{M}_{\rm scal} \,=\, \SO(1,1)\times  \frac{\SO(5,\mathfrak{n})}{\SO(5)\times \SO(\mathfrak{n})}\ ,
\ee
where the first factor is spanned by $\Sigma$ while the second factor is spanned by the scalars in the vector multiplet, which we denote by $\phi^x$, $x=1\ldots,5\mathfrak{n}$.
We indicate the coset representative of the second factor by $\cV =(\cV_M{}^m, \cV_M{}^a)$, where $M=1,\ldots,\mathfrak{n}+5$ labels the fundamental representation of $\SO(5,\mathfrak{n})$. Being an element of $\SO(5,\mathfrak{n})$ this obeys
\begin{equation} \label{eq:vielbein_metric}
 \eta_{MN} = - \cV_M{}^m \cV_N{}^m + \cV_M{}^a \cV_N{}^a \ ,
\end{equation}
where  $\eta_{MN}={\rm diag}(-1,-1,-1,-1,-1,+1,\dots,+1)$ is the flat $\SO(5,\mathfrak{n})$ metric, which is also used to raise and lower the $M,N$ indices (while the $m,n$ and $a,b$ indices are contracted with the $\SO(5)$ and $\SO(\mathfrak{n})$ Kronecker delta, respectively). Alternatively, the coset can be represented by the positive definite scalar metric
\begin{equation}
M_{MN}\ =\ \cV_M{}^m \cV_N{}^m + \cV_M{}^a \cV_N{}^a
\ ,
\end{equation}
which also plays the role of the gauge kinetic matrix for the $(5+\mathfrak{n})$ vector fields $A_{\mu}^{M}=(A_{\mu}^{m}, A_{\mu}^{a})$. The metric on the scalar manifold, which determines the scalar kinetic terms, is
\be\label{scal_metric}
\diff s^2(\mathcal{M}_{\rm scal}) \, =\, 3\Sigma^{-2}\diff \Sigma^2 - \tfrac{1}{8}\diff M_{MN} \diff M^{MN}\ .
\ee

The isometry group of the scalar manifold, $\SO(1,1)\times \SO(5,\mathfrak{n})$, is the global symmetry group of the ungauged supergravity action. In addition, the scalar field space admits a local invariance under $\SO(5)\times \SO(\mathfrak{n})$. The group $\SO(5)$ is promoted to ${\rm Spin(5)}\simeq \USp(4)$ when discussing the couplings to the fermions. It is then convenient to convert the $\SO(5)$ index $m$ of the scalar vielbeine $\cV_M{}^m$ into $\USp(4)$ indices $i,j$ via $\SO(5)$ gamma matrices,
\be\label{vielij_to_vielm}
\cV_M{}^{ij} = \tfrac{1}{2} \cV_M{}^m \Gamma_m^{ij}\ .
\ee
This satisfies $\cV_M{}^{ij}=\cV_M{}^{[ij]}$ and $\Omega_{ij}\cV_M{}^{ij}=0$ and hence transforms in the ${\bf 5}$ of $\USp(4)$. Here $\Omega_{ij}$ is a $4\times4$ real symplectic matrix.

In gauged supergravity a subgroup of the global symmetry group $\SO(1,1)\times \SO(5,\mathfrak{n})$ is promoted to a local gauge symmetry by introducing minimal couplings to the
gauge fields and their supersymmetric counterparts. In this way part of the global symmetry group is broken. When some vector fields transform in non-trivial non-adjoint representations of the gauge group, additional St\"uckelberg-like couplings to antisymmetric rank-two tensor fields may be required in order to ensure closure of the gauge symmetry algebra. Such vector fields can then be gauged away, leaving just massive tensor fields together with the other vectors~\cite{Romans:1985ps,Dall'Agata:2001vb,Schon:2006kz}.

The possible gaugings are classified by the embedding tensor formalism \cite{Nicolai:2000sc,deWit:2002vt,deWit:2004nw}. This introduces
the gauge couplings via
 a spurionic object -- the embedding tensor -- and
 also requires auxiliary fields that consist of a tensor field for each of the original vector fields. In $\mathcal{N}=4$ supergravity, the embedding tensor splits into three different representations of $\SO(1,1)\times \SO(5,\mathfrak{n})$, denoted by $\xi_{M},\xi_{MN}=\xi_{[MN]}$ and $f_{MNP}=f_{ [MNP]}$. Their transformation under $\SO(5,\mathfrak{n})$ follows from the indicated index structure.
With respect to  $\SO(1,1)$, $\xi_{M}$ and $f_{MNP}$ carry  charge $-1/2$, while $\xi_{MN}$ has charge $1$. Supersymmetry of the Lagrangian imposes a set of quadratic constraints on the embedding tensor, whose possible solutions parametrize the different consistent gauged $\mathcal{N}=4$ supergravity theories. In this paper we are interested in theories admitting at least one fully supersymmetric AdS$_5$ vacuum. In \cite{Louis:2014gxa} it was shown that a necessary condition for this is $\xi_{M}=0$. This means that the $\SO(1,1)$ part of the global symmetry is not involved in the gauging and the gauge group is entirely contained in $\SO(5,\mathfrak{n})$. We therefore take $\xi_{M}=0$ from now on. In this case, the quadratic constraints are simply given by
\begin{equation}\label{eq:quadconstr}\begin{aligned}
f_{ R[MN} f_{ PQ]}{}^R\, =&\ 0\ ,
\\
\xi_M{}^Qf_{QNP} \,= &\ 0 \ .
\end{aligned} \end{equation}
The $f_{MNP}$ correspond to structure constants for a (non-Abelian) subgroup of $\SO(5,\mathfrak{n})$, while the $\xi_{MN}$ assign the charges under the $\U(1)$ gauge field $A^0_\mu$.

The embedding tensor determines the gauge covariant derivatives,
\begin{equation}
  \label{s2:DMMN}
  D_{\mu} \, =\, \nabla_{\mu} - A_\mu^M f_M{}^{NP} t_{NP} - A_\mu^0 \,\xi^{NP} t_{NP} \ ,
\end{equation}
where $t_{MN}=t_{[MN]}$ generate $\mathfrak{so}(5,\mathfrak{n})$. It also determines the shift matrices that appear in the fermion supersymmetry variations and  specify the scalar potential.

In the following we abbreviate the contraction of the embedding tensor components $f^{MNP}$ and $\xi^{MN}$ with the coset representatives $\cV_M{}^m$ and $\cV_M{}^a$ by
\begin{align}\label{dressed_emb_tensor}
\hat{f}^{mnp} \,= &\ f^{MNP}{\cV_M{}^m}{\cV_N{}^n}{\cV_P{}^p} \ , \qquad \hat{\xi}^{mn} \,=\, \xi^{MN}{\cV_M{}^m}{\cV_N{}^n} \ , \nn\\
\hat{f}^{mna} \,= &\ f^{MNP}{\cV_M{}^m}{\cV_N{}^n}{\cV_P{}^a} \ , \qquad \hat{\xi}^{ma} \,=\, \xi^{MN}{\cV_M{}^m}{\cV_N{}^a}  \ ,\nn\\
\hat{f}^{mab} \,= &\ f^{MNP}{\cV_M{}^m}{\cV_N{}^a}{\cV_P{}^b} \ , \qquad \,\ \hat{\xi}^{ab} \,=\, \xi^{MN}{\cV_M{}^a}{\cV_N{}^b} \ , \nn\\
\hat{f}^{abc} \,= &\ f^{MNP}{\cV_M{}^a}{\cV_N{}^b}{\cV_P{}^c}  \ .
\end{align}
These ``dressed'' embedding tensor components will always be denoted by a hat symbol. Since they depend on the scalars, generically they vary along domain wall solutions. Also, they appear in the conditions for supersymmetric AdS vacua.

%%%%%%%%%%%%%%%%%
\subsection{Supersymmetric domain walls}
%%%%%%%%%%%%%%%%%

We will be interested in supersymmetric domain wall solutions. The metric is of the form
\be\label{DWmetric}
\diff s^2 \,=\, \rme^{2\mathcal{A}(r)} \diff s^2(\mathbb{R}^{1,3}) + \diff r^2\,,
\ee
where $\mathcal{A}(r)$ is the warp factor which depends only on the radial coordinate $r$.
The one- and two-form supergravity fields vanish, while the scalars have a radial profile, $\Sigma=\Sigma(r)$,  $\phi^x = \phi^x(r)$. In particular, when the solution is AdS$_5$, the scalars are constant and we have $\mathcal{A} = r/\ell$, where $\ell$ is the AdS radius. The latter is related to the cosmological constant, which in our conventions is the same as the critical value of the scalar potential,  $V = -6/\ell^2$.

The supersymmetry conditions for solutions of this form (and with $\xi_M=0$) read \cite{Cassani:2012wc}
\begin{align}
- \mathcal{A}' \gamma_5 \epsilon_i + \iu P_{i}{}^{j} \epsilon_j &= 0 \ , \label{eq:susyeq1}\\[1mm]
\epsilon'_i + \phi^{x}{}' \omega_{x\, i}{}^j \epsilon_j - \tfrac{\iu}{2} P_{i}{}^{j} \gamma_5 \epsilon_j &= 0\ , \label{eq:susyeq2}\\[1mm]
\Sigma' \gamma_5 \epsilon_i +  \iu \Sigma^2\partial_\Sigma P_{i}{}^{j} \epsilon_j &= 0 \ , \label{eq:susyeq3}\\[1mm]
\iu\phi^{x\prime}  v_{x}^{a\,ij}\gamma_5\epsilon_j  - 2 P^{a\,ij} \epsilon_j & = 0
\label{eq:susyeq4} \ ,
\end{align}
where $\epsilon_i$ are the supersymmetry parameters, satisfying the symplectic-Majorana condition $\epsilon_i = \Omega_{ij} C (\bar{\epsilon}^{\,j})^T$.
A prime means derivative with respect to $r$
and $v_x^{am}$ are the vielbeins on the $\frac{\SO(5,\mathfrak{n})}{\SO(5)\times \SO(\mathfrak{n})}$ scalar manifold, defined as
\be
\diff \phi^xv_x^{am}=-(\mathcal{V}^{-1}\diff \mathcal{V})^{am}\ .
\ee
Moreover we introduced the shift matrices
\be\label{gravitino_shift}
P^{ij} = P^{mn}\Gamma_{mn}{}^{ij}\,,\qquad \text{with}\qquad
P^{mn} = -\tfrac{1}{6\sqrt{2}}\, \Sigma^2\, \hat\xi^{mn} + \tfrac{1}{36} \Sigma^{-1} \varepsilon^{mnpqr} \hat{f}_{pqr} \ ,
\ee
where $\varepsilon^{mnpqr}$ is the totally antisymmetric symbol, and
\be\label{gaugino_shift}
P^a{}_{ij} = \tfrac{1}{2\sqrt 2} \Sigma^2\, \hat\xi^{am} \Gamma_{m\,ij} + \tfrac{1}{4}\Sigma^{-1} \hat f^{amn} \Gamma_{mn\,ij}\ .
\ee
The shift matrices also determine the scalar potential as
\be
\label{scalarpot}
 V \,=\, \tfrac{1}{2}P^{a\,ij} P_{a\,ij} + \tfrac{3}{8}\Sigma^2\,\partial_\Sigma P^{ij}\partial_\Sigma P_{ij}
- \tfrac{3}{2}P^{ij}P_{ij}\ .
\ee

The supersymmetry conditions above are obtained by setting to zero the fermion variations given in \cite{Schon:2006kz}.\footnote{The fermionic shifts given in \cite{Schon:2006kz} are related to $P^{ij}$ and $P^{a\,ij}$ appearing here as $A_1^{ij}=\sqrt{\frac{3}{2}} P^{ij}$, $A_2^{ij}= - \sqrt{\frac{ 3}{8}}\,\Sigma\,\partial_\Sigma P^{ij}$, $A_2^{a\,ij} = \frac{1}{\sqrt2} P^{a\,ji}$. For the scalar manifold geometry and the Clifford algebra we use the same conventions as in \cite{Cassani:2012wc}. We have reabsorbed the gauge coupling constant $g$ appearing in \cite{Schon:2006kz} into the embedding tensor.
} Eqs.~\eqref{eq:susyeq1}, \eqref{eq:susyeq2} arise from the gravitino variation, \eqref{eq:susyeq3} arises from the variation of the spin 1/2 fermion in the $\mathcal{N}=4$ gravity multiplet, while \eqref{eq:susyeq4} comes from the gaugino variation. The derivation of~\eqref{eq:susyeq1}, \eqref{eq:susyeq2} assumes that the supersymmetry parameters depend on the coordinate $r$ but are constant on $\mathbb{R}^{1,3}$; this means that we are only describing the Poincar\'e supersymmetries. For generic domain walls these are all the supersymmetries allowed, however in the special case of AdS solutions one also has the conformal supersymmetries, which depend on the coordinates on $\mathbb{R}^{1,3}$. For this reason, the case of AdS solutions will be analyzed separately in the next sections.

As we discuss in detail later, the domain wall supersymmetry conditions imply the existence of a real superpotential function $W$ constructed out of the shift matrix $P^{mn}$, which drives the flow of the warp factor and the scalar fields.
Introducing an index $X=(0,x)$, we can denote the scalars as $\phi^X=(\Sigma,\phi^x)$ and the scalar kinetic matrix as
\be
g_{XY} = \left(\begin{matrix} 3\Sigma^{-2} & 0 \\ 0 & g_{xy} \end{matrix}\right) \ .
\ee
Then the flow equations read
\be\label{floweqs_general}
\mathcal{A}' = W \ ,\qquad \phi^{X}{}' = - 3\, g^{XY}\partial_Y W\ .
\ee
However, this is not the full information encoded into supersymmetry. Indeed, one also finds a set of algebraic constraints restricting the scalar fields that can possibly flow.
{\it After} these constraints are satisfied, the scalar potential  \eqref{scalarpot} can be expressed in terms of the superpotential as
\be\label{VfromW_general}
V \,=\, \tfrac{9}{2}\,g^{XY}\partial_X W\partial_Y W - 6W^2\ .
\ee
This is sufficient to ensure that the Einstein and scalar equations of motion are satisfied \cite{Skenderis:1999mm, DeWolfe:1999cp}. When in particular the superpotential is extremized, $\partial_X W = 0$, we obtain an AdS solution with radius $\ell^{-1} = W$.

The specific form of the superpotential and of the constraints depends on the amount of supersymmetry being preserved and will be discussed in the next sections.

%%%%%%%%%%%%%%%%%%%%%%%
\section{Holographic flows between $\mathcal{N}=2$ SCFTs}
\label{sec:onlyN=4vacua}
%%%%%%%%%%%%%%%%%%%%%%%

In this section, we first review the conditions for fully supersymmetric AdS$_5$ vacua in half-maximal gauged supergravity. Then we show that if there is one such vacuum and the gauge group does not contain any compact part in addition to the $\U(1)\times \SU(2)$ R-symmetry of the vacuum, then the latter is unique, up to moduli.
If on the other hand there is one $\mathcal{N}=4$ vacuum preserving an $\SO(3)$ in addition to the R-symmetry and a certain condition on the gauge coupling constants is satisfied, then we show that there exists a second $\mathcal{N}=4$ AdS vacuum and we construct an explicit flow connecting the two.

%%%%%%%%%%%%%%%%%
\subsection{Review of conditions for $\mathcal{N}=4$ AdS$_5$ vacua}
\label{sec:N=4AdSconditions}
%%%%%%%%%%%%%%%%%

It was shown in \cite{Louis:2015dca} that the necessary and sufficient conditions for five-dimensional half-maximal supergravity to admit a fully supersymmetric AdS$_5$ solution amount to a simple set of constraints on the dressed components of the embedding tensor. In addition to the aforementioned $\xi^{M}=0$, these conditions read:
 \begin{align}
\hat{\xi}^{[mn}\hat{\xi}^{pq]} \,=&\ 0\ ,\label{eq:N=4condA} \\
\hat{\xi}^{ma} \,=&\  0\ , \label{eq:N=4condB}\\
\hat{f}^{mna}\,= &\ 0\ , \label{eq:N=4condC} \\
6\sqrt{2}\, \Sigma^3\,  \hat{\xi}_{mn}\,= &\ -\varepsilon_{mnpqr}  \hat{f}^{pqr} \label{eq:N=4condD}\ ,
\end{align}
where necessarily $\hat\xi^{mn}$ and $\hat f^{mnp}$ are not identically zero.\footnote{Condition \eqref{eq:N=4condD} differs by a factor of $-2$ from the one given in~\cite{Louis:2015dca} because we are including a factor of $1/2$ in the map \eqref{vielij_to_vielm} and when evaluating the shift matrices of~\cite{Schon:2006kz} we are taking $\mathcal{V}^{P m}= -\eta^{PQ}\mathcal{V}_Q{}^m$. See footnote 5 in~\cite{Louis:2015dca}.}
The first condition arises from the gravitino equation while \eqref{eq:N=4condB}--\eqref{eq:N=4condD} are equivalent to $P^{a\,ij}=\partial_\Sigma P^{ij}=0$.
The AdS cosmological constant is read from the scalar potential \eqref{scalarpot} and is
\begin{equation}\label{eq:relccxisq}
V  \,=\,
- \tfrac{3}{2}\,\Sigma^4\,\hat\xi^{mn}\hat\xi_{mn} \ .
\end{equation}

The conditions above imply~\cite{Louis:2015dca} that the theory has gauge group
\begin{equation}
  G \,=\, {\rm U}(1)\times H_{\rm nc} \times   H_{\rm c} \ \subset\ \SO(5,\mathfrak{n}) \ , \label{eq:N=4gaugegroup}
\end{equation}
where $H_{\rm c} \subset \SO(\mathfrak{n})$ is a compact semi-simple subgroup, while $H_{\rm nc}$ is a generically non-compact group admitting $\SO(3)$ as maximal compact subgroup. If $H_{\rm nc}$ is simple, it can be either
$\SO(3)$, $\SO(3,1)$, or ${\rm SL}(3,\mathbb{R})$. When $\hat\xi^{ab}=0$, the product of the $\U(1)$ factor in $G$ with the $\SO(3)$ subgroup of $H_{\rm nc}$ embeds block-diagonally as $\SO(2)\times \SO(3)$ in $\SO(5)$. If $\hat\xi^{ab}\neq 0$, the $\U(1)$ factor is a diagonal subgroup of $\SO(2)\subset \SO(5)$ and  $\SO(2)\subset \SO(\mathfrak{n})$. In the vacuum, the gauge vectors of $\U(1)$ and of $\SO(3) \subset H_{\rm nc}$ are graviphotons, with $\U(1)$ being always gauged by the vector $A^0$, while the gauge vectors of $H_{\rm c}$ and of the non-compact generators of $H_{\rm nc}$ belong to vector multiplets. The non-compact part of $H_{\rm nc}$ is spontaneously broken and the corresponding gauge vectors are massive. Finally, the vectors that are charged under the $\U(1)$ factor of the gauge group are eaten up by antisymmetric rank-two tensor fields via the St\"uckelberg mechanism.
In total, the AdS vacuum is invariant under $\U(1)\times \SU(2) \times H_{\rm c}$. The $\U(1)\times\SU(2)$ corresponds to the R-symmetry of the dual $\mathcal{N}=2$ SCFT, while $H_{\rm c}$ represents the flavor group of that SCFT.

These properties are most easily seen if we perform a global $\SO(1,1)\times \SO(5,\mathfrak{n})$ transformation sending the $\mathcal{N}=4$ critical point to the origin of the scalar manifold, so that $\Sigma =1$ and $(\cV_M{}^m,\cV_M{}^a)$ is the identity element of $\SO(5,\mathfrak{n})$. By further making an $\SO(5)\times \SO(\mathfrak{n})$ transformation, we can choose
\begin{equation}\label{f123xi45}
f^{123} \,=\, g\ ,
\qquad
\xi^{45} \,=\, - \tfrac{1}{\sqrt 2}\,g \ ,
\end{equation}
and the only other non-vanishing components are of the form $f^{1AB},f^{2AB},f^{3AB}, f^{ABC}$ and $\xi^{AB}$, where $A,B,C = 6,7, \dots , \mathfrak{n}+5$. Then $f^{123}$ are $\SU(2)$ structure constants, while the non-vanishing $\xi^{45}$ implies that the vectors $A^4_\mu$, $A^5_\mu$ are eaten up by tensor fields.
Moreover, $f^{1AB},f^{2AB},f^{3AB}$ complete the $\SU(2)$ structure constants to those of $H_{\rm nc}$, while $f^{ABC}$ are the $H_{\rm c}$ structure constants.

From \eqref{eq:relccxisq} we find that the cosmological constant is
\be
V = - \tfrac{3}{2} g^2\ .
\ee

The $\mathcal{N}=4$ vacuum may admit a set of moduli, namely flat directions of the scalar potential along which full supersymmetry is preserved. These are deformations of $\cV_M{}^4$ and $\cV_M{}^5$ such that $\xi^{45}$ is invariant, i.e.
\begin{equation} \label{eq:N=4moduli}
\cV_M{}^4\cV_N{}^5 \xi^{MN} = \xi^{45}\ .
\end{equation}
It was proven in~\cite{Louis:2015dca} that these moduli span the space $\frac{{\rm SU}(1,\mathfrak{m})}{{\rm U}(1) \times {\rm SU}(\mathfrak{m})}$ for some~$\mathfrak{m}$.

%%%%%%%%%%%%%%%%%
\subsection{Uniqueness in the absence of flavor symmetries}
\label{sec:uniquenessN=4_5d}
%%%%%%%%%%%%%%%%%

In the absence of any flavor symmetries $H_{\rm c}$ we can prove that there cannot be two $\mathcal{N}=4$ AdS$_5$ solutions with different values of the cosmological constant. We arrive at this result by showing that in any two such solutions the contractions $\hat\xi_{mn}\hat\xi^{mn}$ and $\hat{f}_{mnp}\hat{f}^{mnp}$ must take the same value. From \eqref{eq:N=4condD} we infer that the $\SO(1,1)$ scalar $\Sigma$ is also unchanged. Then from \eqref{eq:relccxisq} we conclude that the cosmological constant takes the same value in the two solutions.

We first consider the $\xi^{MN}$ components of the embedding tensor, in their dressed version $\hat\xi \equiv \cV^T \xi \cV = {\hat\xi^{mn}\ \hat\xi^{mb} \choose \hat\xi^{an}\ \hat\xi^{ab}}$. The supersymmetry conditions \eqref{eq:N=4condB}, \eqref{eq:N=4condA} and the spectral theory of real, antisymmetric matrices imply that by a local $\SO(5)\times \SO(\mathfrak{n})$ transformation, $\hat\xi$ evaluated on the solution can be put in the canonical block-diagonal form:
\be\label{xicanonical}
\hat\xi  \ = \ {\rm diag}\left(\,0,0,0, \alpha \epsilon, \beta_1 \epsilon,\beta_2\epsilon, \ldots , \beta_p\epsilon,0,\ldots,0 \,\right)\ ,
\ee
where $\epsilon = {\;0\;\; 1 \choose -1\,0}$, while $\pm \iu\alpha$ are the only non-vanishing eigenvalues of $\hat\xi^{mn}$ and $\pm \iu\beta_1$, $\pm \iu\beta_2,\ldots,\pm \iu\beta_p$ are the non-vanishing eigenvalues of $\hat\xi^{ab}$. It is understood that when $\hat\xi^{ab}=0$ there are no $\beta$ eigenvalues. Let us now assume there are two different field configurations corresponding to $\mathcal{N}=4$ AdS$_5$ solutions. The two corresponding vielbeins $\mathcal{V}$ are related by an $\SO(5,\mathfrak{n})$ transformation. However the latter cannot change the eigenvalues of $\hat\xi$, neither can it reshuffle the $\alpha$ eigenvalue with the $\beta$'s, because the former lives in the timelike eigenspace while the latter live in the spacelike eigenspace. It follows that $\hat\xi$ is the same in the two vacua up to $\SO(5)\times \SO(\mathfrak{n})$ transformations. In particular, $\hat\xi_{mn}\hat\xi^{mn} = 2\alpha^2$ is the same in the two vacua.

We now turn to the $f^{MNP}$ components of the embedding tensor. We can assume with no loss of generality that one of the $\mathcal{N}=4$ AdS$_5$ solutions sits at the origin of the scalar manifold.
In an $\SO(5)$ gauge such that \eqref{f123xi45} is true, the other $\mathcal{N}=4$ AdS vacuum must admit an ${\rm SU}(2)\subset H_{\rm nc}$ gauge group with structure constants
$\hat{f}^{123}=\cV_M{}^1\cV_N{}^2\cV_P{}^3 f^{MNP}.$
The choice of an ${\rm SU}(2)$ subgroup inside $H_{\rm nc}$ is described by the coset $H_{\rm nc}/{\rm SU}(2)$. Hence the first three components of the coset representative in the two vacua are related as
\begin{equation} \label{eq:vielbeinN=4} \begin{aligned}
\cV_M{}^1\, = &\ \Lambda_M{}^N  \delta_N{}^1 \ , \\
\cV_M{}^2\, = &\ \Lambda_M{}^N \delta_N{}^2 \ , \\
\cV_M{}^3\, = &\ \Lambda_M{}^N \delta_N{}^3 \ ,
\end{aligned}\end{equation}
with $\Lambda \in H_{\rm nc}$ being given by
\begin{equation} \label{embedding_matrix}
\Lambda_M{}^N \,=\, \left( \begin{aligned} \Lambda_m{}^n && \Lambda_m{}^b \\  \Lambda_a{}^n  && \Lambda_a{}^b \end{aligned} \right)\,=\, {\rm exp} \left( \begin{aligned} 0 && \mu^c f_{cm}{}^b \\  \mu^c f_{ca}{}^n  && 0 \end{aligned} \right) \ ,
\end{equation}
where $(f_c)_{m}{}^b$ are the non-compact generators of $H_{\rm nc}$ and $\mu^c$ are free real parameters.
These transformations $\Lambda$ have been identified in \cite{Louis:2014gxa,Louis:2015dca} as the Goldstone bosons for the spontaneous breaking $H_{\rm nc} \to {\rm SU}(2)$.
Since the $\mathcal{V}$'s in the two AdS$_5$ vacua are related by a gauge transformation, the structure constants $\hat f^{mnp}$ should be preserved. This can easily be seen at first order in $\mu^c$ recalling that \eqref{eq:N=4condC} holds for the vacuum at the origin:
\begin{align}
\hat{f}^{123}\,=&\ \cV_M{}^1\cV_N{}^2\cV_P{}^3 f^{MNP} \,= \, \Lambda_M{}^1\Lambda_N{}^2\Lambda_P{}^3 f^{MNP} \nn\\[2mm]
\,=&\  f^{123} + 3\mu^c f_{ca}{}^{[1} f^{23]a} + \mathcal{O}(\mu^2) \,=\,  f^{123} + \mathcal{O}(\mu^2)\ .
\end{align}
In particular, $\hat{f}_{mnp}\hat{f}^{mnp}$ takes the same value in the two vacua. This concludes our proof.

We remark that a similar argument of uniqueness for fully supersymmetric AdS vacua when $H_c$ is trivial can be derived in $\mathcal{N}=4$ supergravity in dimensions four, six and seven. We provide this in Appendix \ref{app:uniqueness4d}.

%%%%%%%%%%%%%%%%%
\subsection{Two distinct $\mathcal{N}=4$ AdS$_5$ vacua}
\label{sec:twoN=2vacua}
%%%%%%%%%%%%%%%%%

Now let us assume that the $H_{\rm c}$ part of the gauge group is non-trivial.
Since by definition $H_{\rm c}\subseteq \SO(\mathfrak{n})$ and does not contain any $\U(1)$ factor, a non-trivial $H_{\rm c}$ must contain an $\SO(3)_c$ subgroup. As we are going to show, in this case one may have multiple fully supersymmetric vacua by modifying the choice of the $\SO(3)$ subgroup corresponding to the $\SU(2)\times\U(1)$ vacuum R-symmetry within the full gauge group $G$ given in \eqref{eq:N=4gaugegroup}.

 We will assume in the following that the first vacuum is set at the origin of the scalar manifold and is invariant under $H_c$ (hence the dual SCFT has $H_{\rm c}$ flavor symmetry).
In the second vacuum, the $\U(1)$ part of the R-symmetry must also be a diagonal subgroup of $\SO(2)\subset \SO(5)$ and $\SO(2)\subset \SO(\mathfrak{n})$. Since $A^0$ is the gauge vector of that $\U(1)$ globally over scalar field space, this can only be if $\cV_M{}^4$ and $\cV_M{}^5$ differ from their values in the first vacuum by moduli, that is $\hat\xi^{45}=\xi^{45}$, as discussed in Section~\ref{sec:N=4AdSconditions}. Therefore the two vacua are only distinguished by the values of $\cV_M{}^m$ for $m=1,2,3$.
The condition \eqref{eq:N=4condC} then means that in the second vacuum we find an $\SO(3)_2$ subgroup of $G$ that is gauged by $\hat{A}^m=A^M\mathcal{V}_M{}^m, m=1,2,3$. Most generally this subgroup can be a subgroup $\SO(3)_2 \subset \SO(3)_1 \times \SO(3)_c$, where $\SO(3)_1$ is part of the R-symmetry in the original vacuum, while $\SO(3)_c \subset H_c$.
We can use $\SO(5,\mathfrak{n})$ rotations to choose this $\SO(3)_c$ group to be in the $M=6,7,8$ directions at the origin. We will denote the $\SO(3)_c$ structure constants by
\begin{equation} \label{eq:N=2gauge_comp}
 f^{678} = g \lambda^{-1} \ ,
\end{equation}
where $\lambda$ is a real constant, while as before we will take
\begin{equation}\label{f123xi45when_two_vacua}
f^{123} \,=\, g\ ,
\qquad
\xi^{45} \,=\, - \tfrac{1}{\sqrt2}\,g
\end{equation}
for the gauge coupling constant of $\SO(3)_1 \times \U(1)$. The gauge fields of $\SO(3)_1$ are thus $A^{1,2,3}$, those of $\SO(3)_c$ are $A^{6,7,8}$, while $A^{4,5}$ are eaten up by tensor fields since they are charged under the $\U(1)$ gauged by $A^0$.

As already seen before, the embedding tensor above leads to an $\mathcal{N}=4$ vacuum at the origin of the scalar manifold, with cosmological constant $V = - \tfrac{3}{2} g^2$.
We can also assume that in the second vacuum the coset representative $\cV_M{}^m$ has for $m=1,2,3$ only components in the $M=m$ and $M= m+5$ directions.
We then parametrize the coset representative as
\be
\mathcal{V}=\rme^{-2\phi_1 t_{16}-2\phi_2 t_{27}-2\phi_3 t_{38}}\ ,
\ee
where $(t_{MN})_P{}^Q=\delta^Q_{[M}\eta^{\,}_{N]P}$ are the generators of $\mathfrak{so}(5,\mathfrak{n})$ in the fundamental representation. More explicitly, its non-trivial part is
\begin{equation}\label{eq:N=4ansatz}
\cal{V} \,=\, \left(\begin{matrix} \cosh\phi_1 & 0 & 0 & 0 & 0 & -\sinh\phi_1 & 0 & 0 & \ldots\\
0 & \cosh\phi_2 & 0 & 0 & 0 & 0 & -\sinh\phi_2 & 0 & \ldots\\
0 & 0 & \cosh\phi_3 & 0 & 0 & 0 & 0 & -\sinh\phi_3 & \ldots\\
 0 & 0 & 0 & 1 & 0 & 0 & 0 & 0 & \ldots\\
0 & 0 & 0 & 0 & 1 & 0 & 0 & 0 & \ldots\\
 -\sinh\phi_1 & 0 & 0 & 0 & 0 & \cosh\phi_1 & 0 & 0 & \ldots\\
0 & -\sinh\phi_2 & 0 & 0 & 0 & 0 & \cosh\phi_2 & 0 & \ldots\\
 0 & 0 & -\sinh\phi_3 & 0 & 0 & 0 & 0 & \cosh\phi_3 & \ldots\\
 \vdots & \vdots & \vdots & \vdots & \vdots & \vdots & \vdots & \vdots & \ddots\\
 \end{matrix}\right).
\end{equation}
With the choice above for the embedding tensor and for the scalar fields, the only non-trivial $\mathcal{N}=4$ condition on the scalars $\phi_m$ is given by \eqref{eq:N=4condC}, which leads to
\begin{equation}
 \tanh\phi_m \tanh\phi_n = \lambda \tanh\phi_p\ ,
\end{equation}
with $(m,n,p)$ cyclic permutations of $(1,2,3)$. Apart for the trivial solution $\phi_m=0$, these equations have the solution $\phi_1=\phi_2=\phi_3=\phi$ (or $\phi_1=-\phi_2=-\phi_3=\phi$, etc.), with
\begin{equation}
\tanh\phi = \lambda \ .
\end{equation}
This implies that a second vacuum can only exist for
\begin{equation}\label{eq:secondN=4vac_cond}
|\lambda| < 1 \ .
\end{equation}
In that vacuum, we find that the coupling constant of $\SO(3)_2$ is
\begin{equation}\label{str_const_IR_N=4vacuum}
  \hat{f}^{123} = g \left( 1- \lambda^2 \right)^{-1/2} \ .
\end{equation}
Using this and the fact that $\hat\xi^{45}=\xi^{45}= - \tfrac{1}{\sqrt2}\,g$, we find from \eqref{eq:N=4condD} that the scalar $\Sigma$ is determined as
\begin{equation}
  \Sigma = \left(1-  \lambda^2\right)^{-1/6} \ .
\end{equation}
Then \eqref{eq:relccxisq} gives for the cosmological constant
\begin{equation}
  V = - \,\tfrac{3}{2}\, g^2 \left(1-\lambda^2\right)^{-2/3} \ .
\end{equation}

In order to identify which gauge symmetries are spontaneously broken we study the covariant derivative of the scalar fields around the second vacuum.
Starting from \eqref{s2:DMMN}, one can see that in general the scalar covariant derivative reads
\be\label{ScalarCovDer_general}
D\phi^{am} = \diff \phi^{am} - \hat{\xi}^{am} A^0 + \hat{f}^{amn}\hat{A}_n - \hat{f}^{amb} \hat{A}_b\ ,
\ee
where $\hat{A}^n=A^P\mathcal{V}_P{}^n$, $\hat{A}^a=A^P\mathcal{V}_P{}^a$ are dressed vectors and we have defined $\diff \phi^{am} \equiv v_x^{am}\diff \phi^x$.

We expand the covariant derivative at first order in the field fluctuations around the second vacuum.
In particular the constants
\begin{equation}
\hat f^{178} = \hat f^{286} =\hat f^{367} = - g \left(1-\lambda^2\right)^{-1/2}
 \, ,
\end{equation}
are non-zero and lead to
\begin{align}
 D (\phi^{17}- \phi^{26}) &= \diff (\phi^{17}- \phi^{26}) + 2 \hat f^{178} \hat A^8
 \nn\\
 &= \diff (\phi^{17}- \phi^{26}) - 2g \left(1-\lambda^2\right)^{-1}\left(A^8 -\lambda A^3 \right) \ ,
\end{align}
while $\phi^{17}+ \phi^{26}$ remains uncharged (here $6,7,8$ denote the values taken by the $a$ index). One also has similar expressions for simultaneous cyclic permutations of the indices $m = (123)$ and $a,b = (678)$. It follows that $\hat A^a = \left(1-\lambda^2\right)^{-1/2}\left(A^a -\lambda A^{a-5} \right)$, with $a=6,7,8$, are all massive, and the gauge group $\SO(3)_1 \times \SO(3)_c$ is indeed broken to the diagonal subgroup $\SO(3)_2$ with structure constant \eqref{str_const_IR_N=4vacuum}, gauged by $\hat{A}^m = (1-\lambda^2)^{-1/2}(A^m - \lambda A^{5+m})$, for $m=1,2,3$.
If moreover $\SO(3)_c$ is part of a larger gauge group $H_c$, and there are other generators of $H_c$ that do not commute with $\SO(3)_c$, then the constants $f^{MNP}, M=6,7,8$ and $N,P >8$ are non-zero. In the second vacuum this leads to non-vanishing structure constants given by $\hat f^{mab} = \sinh \phi f^{(M=m+5)(a=N)(b=P)}$ that give a mass to the gauge vectors corresponding to those symmetries. That means that  $\SO(3)_1 \times H_c$ is spontaneously broken to the product of $\SO(3)_2$ with the maximal commutant of $\SO(3)_c$ in $H_c\,$.

We emphasize that by the procedure above we find a possible second $\mathcal{N}=4$ vacuum for every inequivalent embedding of $\SO(3)_c$ into $H_c$ such that the condition \eqref{eq:secondN=4vac_cond} holds.

In Section~\ref{sec:FlowN=4toN=4} we present a domain wall solution between the two $\mathcal{N}=4$ vacua above and discuss its holographic interpretation.

%%%%%%%%%%%%%%%%%
\subsection{Conditions for flows with eight Poincar\'e supercharges}\label{sec:HalfBPSflows}
%%%%%%%%%%%%%%%%%

Domain wall solutions preserving eight of the sixteen supercharges were only partially discussed in \cite{Cassani:2012wc}. Here we provide their complete characterization (when $\xi_M=0$), which to the best of our knowledge has not appeared in the literature before.\footnote{The analysis is also similar in spirit to the one performed in $\mathcal{N}=2$ supergravity in \cite{Ceresole:2001wi}.}

Starting from the gravitino shift matrix $P$ defined in \eqref{gravitino_shift}, we introduce the superpotential
\begin{equation}\label{eq:N=2superpotential}
W = \sqrt{2\,P_{mn}P^{mn}} \,.
\end{equation}
Then the supersymmetry conditions are equivalent to the flow equations
\begin{align}
& \mathcal{A}'\, =\, W \,,  \label{flow_warp}\\
& \Sigma'\,\, = - \Sigma^2 \partial_\Sigma W  \,,\label{flow_sigma} \\
&\phi^{x}{}' = - 3\, g^{xy}\partial_y W \,, \label{flow_phi}
\end{align}
together with the constraints
\begin{align}
& P^{[mn}P^{pq]} = 0 \label{HalfBPSconstr1}\,,\\
& \partial_\Sigma \left(W^{-1}P^{mn}\right) = 0 \label{HalfBPSconstr2}\,,\\
& \hat f^{amn}P_{mn} = 0 \label{HalfBPSconstr3} \,,\\
& \tfrac{1}{4\sqrt{2}}\Sigma^3 \varepsilon^{mn}{}_{pqr} P^{pq}\hat\xi^{ra} = P_p{}^{[m} \hat f^{n]pa} \label{HalfBPSconstr4}\,.
\end{align}
When these constraints are satisfied, the scalar potential \eqref{scalarpot} can be written in terms of the superpotential as
\be\label{VfromW}
V \,=\, \tfrac{9}{2}\,g^{xy}\partial_x W\partial_y W + \tfrac{3}{2}\,\Sigma^2 (\partial_\Sigma W)^2 - 6W^2\ .
\ee
Clearly, the flow equations and the form of the potential agree with \eqref{floweqs_general}, \eqref{VfromW_general}.

One can show that if the constraints \eqref{HalfBPSconstr1}--\eqref{HalfBPSconstr4} are satisfied and the superpotential $W$ is extremized, then the $\mathcal{N}=4$ AdS conditions of Section \ref{sec:N=4AdSconditions} are recovered. In other words, the fixed points of flows preserving eight supercharges are $\mathcal{N}=4$ AdS solutions. The converse implication is of course also true, as an $\mathcal{N}=4$ AdS$_5$ solution can be seen as a domain wall preserving eight Poincar\'e supercharges and having constant scalars.

%%%%%%%%%%%%%
\subsubsection*{Proof}
%%%%%%%%%%%%%

Let us prove the supersymmetric flow equations above.
We start from the gravitino equation \eqref{eq:susyeq1}. Multiplying by $P$ we obtain
\be
 P^{mn} P^{pq}(\Gamma_{mnpq})_i{}^j  \epsilon_j = \left[2 P_{mn} P^{mn} - \left(\mathcal{A}'\right)^2 \right] \epsilon_i \,.
\ee
In order to solve this equation while preserving eight degrees of freedom in the supersymmetry parameter $\epsilon_i$, we need the two sides to vanish separately~\cite{Cassani:2012wc}. In this way we obtain the constraint \eqref{HalfBPSconstr1} and the evolution equation \eqref{flow_warp} for the warp factor, where $W$ is defined as in \eqref{eq:N=2superpotential}.
Since now
\be
P_{i}{}^{k}P_{k}{}^{j} = - W^2 \,\delta_i{}^j\ ,
\ee
we can write
\begin{equation} \label{decompose_P}
P  = W I \ ,
\end{equation}
so that $I^2 = -1$ is an almost complex structure.
Then the gravitino equation \eqref{eq:susyeq1} takes the form of the projector
\begin{equation} \label{eq:N=2projection}
I_i{}^j \epsilon_j + \iu \gamma_5 \epsilon_i =0\ ,
\end{equation}
which precisely reduces the number of independent components in $\epsilon_i$ by half.

Using the relations just obtained, the differential equation \eqref{eq:susyeq2} for the spinor is solved~by
\begin{equation}
\epsilon_i = \e^{\mathcal{A}/2} \hat \epsilon_i \ ,
\end{equation}
where $\hat \epsilon_i$ is a covariantly constant spinor on $\mathbb{R}^{1,3}$ (with the covariant derivative including the $\USp(4)$ connection).

We now pass to the supersymmetry condition \eqref{eq:susyeq3}. Since it has to hold for any spinor satisfying the projector \eqref{eq:N=2projection}, it must be that
\be
\left(\Sigma' \gamma_5 \delta_i{}^j +  \iu \Sigma^2\partial_\Sigma P_{i}{}^{j} \right)\left(\gamma_5\delta_j{}^k +\iu  I_j{}^k \right) = 0 \ ,
\ee
which is equivalent to
\be
\Sigma'  \delta_i{}^k - \Sigma^2 \partial_\Sigma P_i{}^j I_{j}{}^k = 0\,
\ee
because the terms linear in $\gamma_5$ cannot compensate the others and thus have to vanish separately. Using \eqref{decompose_P} and noting that $I^2=-1$  implies ${\rm Tr}(I \partial_\Sigma I)=0$, gives the flow equation \eqref{flow_sigma} for $\Sigma$, together with constraint \eqref{HalfBPSconstr2}.

It remains to discuss the supersymmetry equation \eqref{eq:susyeq4}.
The same argument used to manipulate equation \eqref{eq:susyeq3} allows to infer that \eqref{eq:susyeq4} together with the projection \eqref{eq:N=2projection} is equivalent to
\be
 \phi^{x\prime}  v_x^{a\,ij} \, =\, 2 P^{a\,ik} I_{k}{}^j \,.
\ee
Separating the terms transforming in different irreducible representations of $\USp(4)$, we get
\begin{align}
P^{a\,ij} I_{ij} \,&=\, 0\ ,\nn\\
P^{a}{}_{(i}{}^k I_{j)k} \,&=\, 0\ ,\nn\\
\tfrac{1}{2}v^{a}_{y\,ij} P^{a\,ik} I_{k}{}^j \,&=\, g_{yx} \phi^{x}{}'\,,
\end{align}
where to obtain the last equation we used $v_x^{a\,ij}v_y^a{}_{ij}=4g_{xy}$.
Recalling the definition of the gaugino shift matrix \eqref{gaugino_shift}, the first and the second are easily seen to correspond to constraints \eqref{HalfBPSconstr3} and  \eqref{HalfBPSconstr4}, respectively. The third instead gives the flow equation \eqref{flow_phi}, because
\be\label{last_step}
\tfrac{1}{2} v^{a}_{y\,ij} P^{a\,ik} I_k{}^j = \,-3\, \partial_y W\,.
\ee
This can be seen by an explicit computation: evaluating the derivative of \eqref{eq:N=2superpotential} one finds
\be
-3 \,\partial_y W =  v_y^{am} \left( \sqrt{2}\, \Sigma^2 \hat{\xi}^{an}I_{nm} + \tfrac{1}{2}\Sigma^{-1}\varepsilon_{mnpqr}I^{np}\hat{f}^{qra}\right),
\ee
where we used $D_x \mathcal{V}_M{}^m = - \mathcal{V}_M{}^a v_x^{am}$.
Exactly the same expression is obtained by evaluating $\tfrac{1}{2} v^{a}_{y\,ij} P^{a\,ik} I_k{}^j$. This concludes our proof.

%%%%%%%%%%%%%%%%%
\subsection{Flow between two $\mathcal{N}=4$ AdS$_5$ vacua and its holographic dual}
\label{sec:FlowN=4toN=4}
%%%%%%%%%%%%%%%%%

We now construct a flow connecting the two $\mathcal{N}=4$ AdS$_5$ vacua discussed in Section~\ref{sec:twoN=2vacua}.  This should correspond to a holographic RG flow connecting two $\mathcal{N}=2$ four-dimensional SCFTs.
We preserve all the eight Poincar\'e supersymmetries along this flow and these get enhanced to sixteen at the AdS fixed points by the eight additional conformal supercharges.

We again use the local symmetry on the scalar manifold to choose the relevant components of the embedding tensor as in \eqref{eq:N=2gauge_comp}, \eqref{f123xi45when_two_vacua}. We see from the solution for the second $\mathcal{N}=4$ vacuum that besides $\Sigma$ the only flowing scalars should be $\phi_1, \phi_2, \phi_3$ in the parametrization~\eqref{eq:N=4ansatz} of the coset representative.
Since we do not want to break the diagonal $\SO(3)_2$ symmetry along the flow, we set the three scalars equal to each other, $\phi_1=\phi_2=\phi_3=\phi$.
 We can then construct the shift matrix \eqref{gravitino_shift} and the superpotential \eqref{eq:N=2superpotential}. We obtain:
\be
P^{mn}= W \delta^{[m}_4\delta^{n]}_5\ ,
\ee
with the superpotential being
\begin{equation}\label{superpot_two_N=4_vacua}
W \,=\, \tfrac{1}{6}g\, \Sigma^{2} +  \tfrac{1}{3} g\,\Sigma^{-1} \left(  \cosh^3\phi -  \lambda^{-1} \sinh^3\phi \right) \ ,
\end{equation}
where we are assuming $g> 0$ for simplicity.\footnote{Strictly speaking, formula \eqref{eq:N=2superpotential} for the superpotential yields the absolute value of the right hand side of~\eqref{superpot_two_N=4_vacua}. However assuming $g>0$ we see that both in the first vacuum $(\Sigma=1,\phi=0)$ and in the second vacuum the right hand side of~\eqref{superpot_two_N=4_vacua} is positive; we can thus remove the absolute value.}
It is easy to check that the constraints \eqref{HalfBPSconstr1}--\eqref{HalfBPSconstr4} are satisfied with no further assumptions.

The metric on the subspace spanned by the two scalars is computed from \eqref{scal_metric} and reads
\be
\diff s^2  =\, 3\,\Sigma^{-2}\diff \Sigma^2 + 3\,\diff\phi^2 \, .
\ee
The scalar potential is
\begin{align}\label{scal_pot_two_N=4_vacua}
V & = \tfrac{1}{2}g^2\Sigma^{-2}\left[\cosh^4\phi\left(\cosh(2\phi) -2\right)  -4\lambda^{-1} \cosh^3\phi\sinh^3\phi  +  \lambda^{-2}\sinh^4\phi\left(\cosh(2\phi) +2\right) \right]    \nn\\[1mm]
& \,\quad -g^2\Sigma\left( \cosh^3\phi - \lambda^{-1} \sinh^3\phi \right)\, .
\end{align}
The superpotential and the scalar potential are related as in \eqref{VfromW_general}, namely
\begin{equation}
V = \tfrac{3}{2}\,\Sigma^2 (\partial_{\Sigma}W)^2+\tfrac{3}{2} (\partial_{\phi}W)^2- 6 W^2\, .
\end{equation}

Imposing extremization of the superpotential, $\partial_{\phi}W = \partial_{\Sigma}W = 0$, one recovers the two fully supersymmetric AdS vacua, that is the one at the origin,
\begin{equation}\label{eq:UVAdS}
\Sigma=1\,, \qquad \phi = 0\,, \qquad W = \tfrac{g}{2}\,,\qquad V = -\tfrac{3}{2}g^2\,,
\end{equation}
and the one at non-trivial values of the scalar fields,
\begin{equation}\label{eq:2ndAdS}
\Sigma=\Sigma_*=\left(1-\lambda^2\right)^{-1/6}\,, \qquad \phi=\phi_* = \tfrac{1}{2}\log\tfrac{1+\lambda}{1-\lambda}\,, \qquad W = \tfrac{g}{2}\Sigma_*^2\,,\qquad V = -\tfrac{3}{2}g^2\Sigma_*^4\,.
\end{equation}
We recall that we should impose $\lambda<1$ in order to have a well-defined vacuum.

It is easy to compute the masses of the scalar fields at these two vacua. They are given by the eigenvalues of the matrix $g^{XY}\partial_{X}\partial_{Y}V$ where $g^{XY}$ is the inverse of the scalar metric.
It is useful to compute the dimensionless scalar mass, i.e.\ the combination $m^2\ell^2$ where $\ell$ is the scale of AdS.
At the UV vacuum one finds
\begin{equation}\label{eq:UVdim}
m^2_{\Sigma}\ell^2 =m^2_{\phi}\ell^2 = -4\;.
\end{equation}
At the IR vacuum one has
\begin{equation}\label{eq:IRdim}
m^2_{\Sigma}\ell^2 = -4 \,,\qquad m^2_{\phi}\ell^2 = 12\;.
\end{equation}

We can now employ the holographic identity $m^2\ell^2 = \Delta(\Delta-4)$ to extract the conformal dimensions of the operators dual to the two scalars at the UV and IR AdS vacua. At the UV vacuum we find that both scalars are dual to operators of dimension $\Delta=2$. In the IR vacuum $\Sigma$ is still dual to an operator of dimension $\Delta_{\Sigma}=2$ and is thus relevant, however the operator dual to $\phi$ is irrelevant and has dimension $\Delta_{\phi}=6$.

Notice that in an $\mathcal{N}=2$ SCFT the energy-momentum multiplet contains the $\SU(2)\times \U(1)$ R-current as well as a real operator of dimension $2$ (see for example page 18 in \cite{Dolan:2002zh}). We thus find that the conformal dimensions computed in \eqref{eq:UVdim} and \eqref{eq:IRdim} are consistent with identifying the scalar $\Sigma$ as the gravitational dual to the operator of dimension $2$ in the energy-momentum multiplet. This is also consistent with the supergravity analysis since $\Sigma$ sits in the gravity multiplet of five-dimensional half-maximal supergravity. Through similar reasoning one finds that the operator dual to the scalar $\phi$ is the bottom component in the UV $\SO(3)$ flavor current multiplet. This operator is sometimes referred to as momentum map operator. It transforms as a triplet of both the R-symmetry and the flavor SO(3)'s and we are giving a vev to the component invariant under the diagonal SO(3) subgroup.

The value of the cosmological constants at the two AdS vacua in \eqref{eq:UVdim} and \eqref{eq:IRdim} determines the ratio of the central charges of the dual SCFTs, see for example \cite{Freedman:1999gp}. We find
\begin{equation}\label{ratioN=2case}
\frac{c_{\rm IR}}{c_{\rm UV}} = \left(\frac{V_{\rm IR}}{V_{\rm UV}}\right)^{-3/2} =\, 1-\lambda^2\,.
\end{equation}
Since $\lambda^2<1$ this result is compatible with the $a$-theorem.
Notice that this is also the same ratio as $(g_{\rm IR}/g)^{-2}$, where $g_{\rm IR}$ is the gauge coupling of the IR R-symmetry, given in \eqref{str_const_IR_N=4vacuum}.

The flow equations generated by the superpotential \eqref{superpot_two_N=4_vacua} via \eqref{flow_warp}--\eqref{flow_phi} read
\begin{align}\label{eq:floweqs_N=4toN=4}
\Sigma' \,& =\,  - \,\tfrac{1}{3}\,g \Sigma^3 + \tfrac{1}{3}\, g\left(  \cosh^3\phi - \lambda^{-1} \sinh^3\phi\right) \ , \nn\\[1mm]
\phi' \,& =\, -  g\,\Sigma^{-1} \sinh\phi \cosh\phi \left( \cosh\phi - \lambda^{-1} \sinh\phi \right) \ ,\nn\\[1mm]
\mathcal{A}' &= W\,.
\end{align}
It is possible to solve analytically for $\Sigma$ and $\mathcal{A}$ as a function of $\phi$. After a short calculation one finds that the solution for $\Sigma$ is
\begin{equation}\label{eq:Sigmaphisol}
\Sigma(\phi) =  \frac{\left(\cosh\phi -\lambda^{-1}\sinh\phi\right)^{1/3}}{\left(\cosh(2\phi)+ c_1 \sinh(2\phi)\right)^{1/3}}\ ,
\end{equation}
where $c_1$ is an integration constant. In order for the solution to reach the IR AdS vacuum in \eqref{eq:2ndAdS} we should fix $c_1 = -\frac{1}{2}\left(\lambda+\lambda^{-1}\right)$.
In a similar way one can find the following solution for the warp factor,
\begin{equation}
\mathcal{A}(\phi) = \tfrac{1}{6}\log\left[\frac{(\sinh\phi-\lambda^{-1}\cosh\phi)(\cosh\phi-\lambda^{-1}\sinh\phi)^3}{\sinh^3(2\phi)}\right] +c_2 \ ,
\end{equation}
where $c_2$ is a trivial integration constant that can be set to any desired value by shifting the radial coordinate $r$. The asymptotic behavior of $\mathcal{A}$ close to the two AdS vacua is
\begin{equation}
\mathcal{A}_{{\rm UV}} \approx -\tfrac{1}{2}\log\phi\,,\qquad \mathcal{A}_{{\rm IR}} \approx \tfrac{1}{2}\log(\phi-\phi_*)\,.
\end{equation}
This is the expected divergent behavior of the metric function close to the two AdS vacua.

\begin{figure}[h]
\centering
\includegraphics[width=7cm]{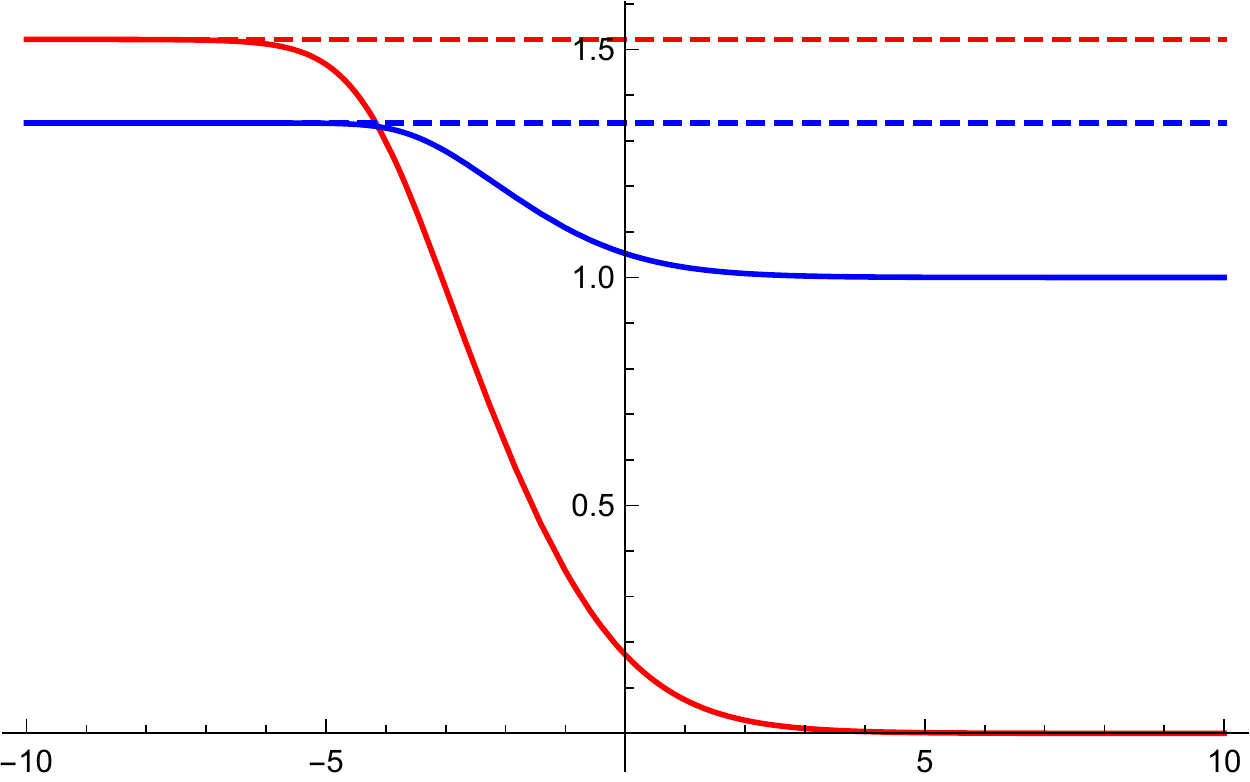} \qquad\qquad \includegraphics[width=7cm]{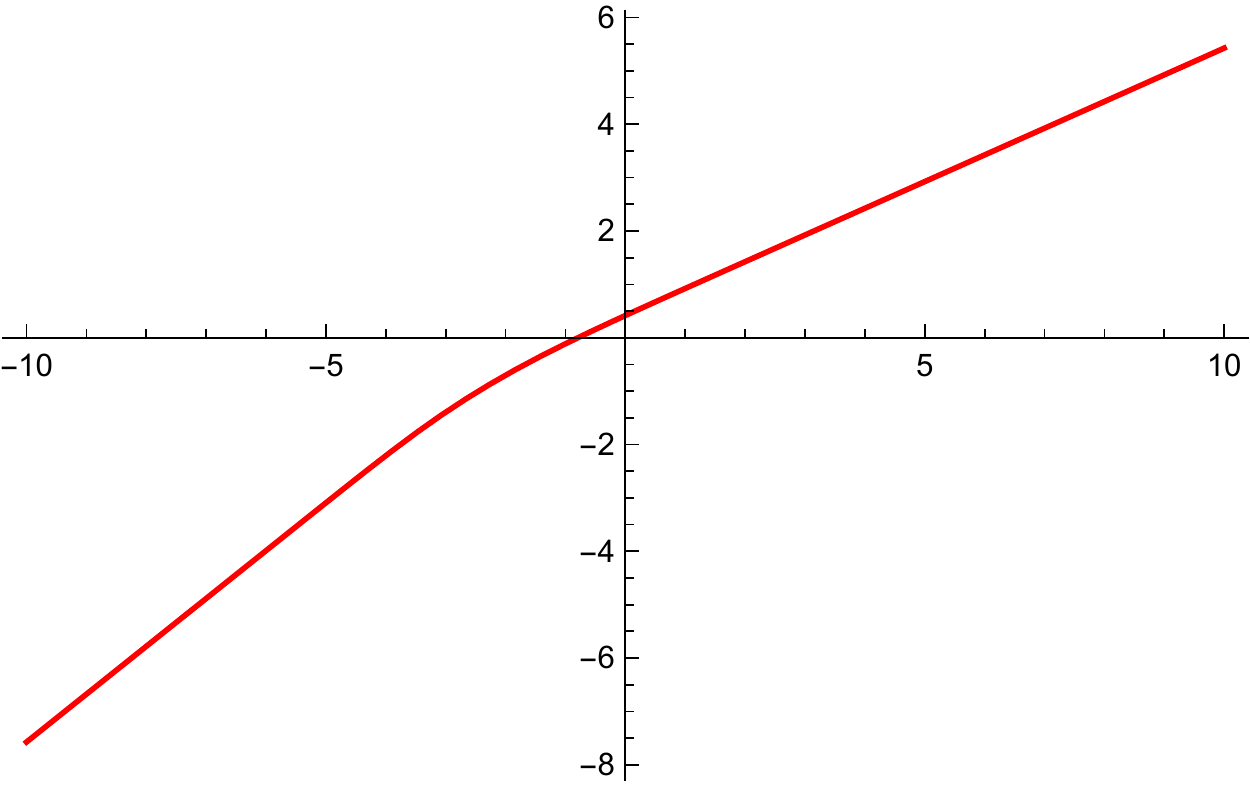}
\caption{\label{fig:Plots} \textit{Left}: Numerical solutions for $\phi(r)$ (red solid line) and $\Sigma(r)$ (blue solid line). The dashed red/blue lines are the values for the scalars at the IR vacuum in \eqref{eq:2ndAdS}. We have fixed $g=1$ and $\lambda^{-1}=1.1$. \textit{Right}: Numerical solution for $\mathcal{A}(r)$ for the same values of $g$ and $\lambda$. The IR/UV is at large negative/positive values of $r$. The function $\mathcal{A}(r)$ is linear in these regions and the scalars attain their fixed point values as expected from \eqref{eq:UVAdS} and \eqref{eq:2ndAdS}}
\end{figure}

One can plug the analytic solution for $\Sigma(\phi)$ back into the second equation in \eqref{eq:floweqs_N=4toN=4} and solve for the function $\phi(r)$ in quadratures. Then one can use this solutions to find also the functions $\Sigma(r)$ and $\mathcal{A}(r)$. We were not able to solve for $\phi(r)$ analytically, however a typical numerical plot for the scalars and metric function is not hard to generate, see Figure~\ref{fig:Plots}.
%%%%%%%%%%
%%%%%%%%%%

It is also instructive to analyze the flow close to the UV AdS vacuum in order to understand what drives it. We can linearize the flow equations in \eqref{eq:floweqs_N=4toN=4} around the vacuum in \eqref{eq:UVAdS} to find
\begin{equation}
\phi' \approx -g \phi\,, \qquad \Sigma' \approx -g(\Sigma-1)\;, \qquad \mathcal{A}' \approx \tfrac{g}{2}\;.
\end{equation}
Using that the AdS scale is $\ell=2/g$ we find the approximate solution
\begin{equation}\label{eq:UVradexp}
\phi \approx v_{\phi} \,\rme^{-2r/\ell}\,, \qquad \Sigma \approx1+ v_{\Sigma}\, \rme^{-2r/\ell}\,, \qquad \mathcal{A} \approx \tfrac{r}{\ell}\,.
\end{equation}
Since the scalars $\phi$ and $\Sigma$ are dual to operators of dimension $2$ in the SCFT we can conclude that the RG flow is driven by vacuum expectation values for these two operators. If there were explicit sources for the operators the approximate UV solution should have had an $r \rme^{-2r/\ell}$ term in the asymptotic expansion. This is clearly absent in our setup.

Expanding the explicit analytic solution in \eqref{eq:Sigmaphisol} around the UV AdS vacuum at $\phi \approx 0$ we find
\begin{equation}
\Sigma \approx 1+ \tfrac{\lambda}{3} \,\phi + \ldots\,.
\end{equation}
We thus conclude that the constants $v_\Sigma$ and $v_\phi$ in \eqref{eq:UVradexp} are related by
\be v_{\Sigma} = \tfrac{\lambda}{3}\, v_{\phi}\ .
\ee
It would be interesting to understand field-theoretically the corresponding relation between the operator vevs.

%%%%%%%%%%%%%%%%%%%%%%%%%
\section{Field theory derivation of ratio between central charges}
\label{sec:QFT}
%%%%%%%%%%%%%%%%%%%%%%%%%

We pause here our supergravity analysis and present a field theory explanation for the ratio of UV and IR central charges of $\mathcal{N}=2$ SCFT's found holographically in \eqref{ratioN=2case}.

%%%%%%%%%%%%%%%%%%%%%%%%%%%%%%
\subsection{Anomalies in four-dimensional $\mathcal{N}=2$ SCFTs}
\label{sec:anomalies}
%%%%%%%%%%%%%%%%%%%%%%%%%%%%%%

The R-symmetry of four-dimensional $\mathcal{N}=1$ SCFTs is $\U(1)_{R_{\mathcal{N}=1}}$. The cubic and linear 't~Hooft anomalies are\footnote{The $\text{Tr}$ symbol in all equations below should be understood formally. In the presence of a Lagrangian it indicates a trace over the charges of the chiral fermions in the theory.}
\begin{equation}
\text{Tr}(R_{\mathcal{N}=1}^3) \qquad \text{and}\qquad \text{Tr}(R_{\mathcal{N}=1})\,.
\end{equation}
Via $\mathcal{N}=1$ supersymmetric Ward identities these anomalies are related to the conformal anomalies by the well-known relations \cite{Anselmi:1997am}
\begin{equation}\label{eq:acdef}
a = \tfrac{9}{32}\text{Tr}(R_{\mathcal{N}=1}^3) - \tfrac{3}{32}\text{Tr}(R_{\mathcal{N}=1})\,, \qquad c = \tfrac{9}{32}\text{Tr}(R_{\mathcal{N}=1}^3) - \tfrac{5}{32}\text{Tr}(R_{\mathcal{N}=1})\,.
\end{equation}

For four-dimensional $\mathcal{N}=2$ SCFTs the R-symmetry is $\SU(2)_R\times \U(1)_{R_{\mathcal{N}=2}}$. The generators of $\SU(2)_R$ are denoted by $I_a$, $a=1,2,3$.\footnote{The indices $a,b$ used in the present field theory section are unrelated to the $\SO(\mathfrak{n})$ indices used in the rest of the paper.} There is a unique $\mathcal{N}=1$ superconformal subalgebra of the $\mathcal{N}=2$ superconformal algebra. This fixes how the $\U(1)_{R_{\mathcal{N}=1}}$ is embedded into the Cartan of the $\SU(2)_R\times \U(1)_{R_{\mathcal{N}=2}}$ R-symmetry, see for example \cite{Aharony:2007dj,Tachikawa:2009tt},
\begin{equation}\label{RNeq1UV}
R_{\mathcal{N}=1} = \tfrac{1}{3} R_{\mathcal{N}=2} +\tfrac{4}{3}I_{3}\ ,
\end{equation}
and it is this $R_{\mathcal{N}=1}$ that is used to compute the conformal anomalies via \eqref{eq:acdef}.
Continuous flavor symmetries in four-dimensional $\mathcal{N}=2$ SCFTs are characterized by a flavor central charge $k_F$ given by the 't~Hooft anomaly (see Eq. (2.6) of \cite{Aharony:2007dj})
\begin{equation}\label{eq:kGdef}
k_F\, \delta_{ab}= - 2 \text{Tr}(R_{\mathcal{N}=2}T_aT_b)\,,
\end{equation}
where $T_a$ are the generators of the flavor group.

%%%%%%%%%%%%%%%%%%%%%%%%%%%%%%
\subsection{RG flow between $\mathcal{N}=2$ SCFTs}
\label{sec:flow}
%%%%%%%%%%%%%%%%%%%%%%%%%%%%%%

We are interested in an RG flow which connects two distinct four-dimensional $\mathcal{N}=2$ SCFTs. In parallel with the supergravity setup, assume that the UV SCFT has $\SU(2)_R\times \U(1)_{R_{\mathcal{N}=2}}$ and an $\SU(2)_F$ flavor symmetry.\footnote{This analysis can be generalized to a more general flavor symmetry group. In that case the discussion below applies to an $\SU(2)$ subgroup of the flavor group.} The generators of the flavor symmetry algebra in the UV will be denoted by $T_a$. In the IR SCFT the symmetry is $\widetilde{\SU(2)}_R\times \U(1)_{R_{\mathcal{N}=2}}$ where $\widetilde{\SU(2)}_R$ is the diagonal subgroup of $\SU(2)_R \times \SU(2)_F$. The UV conformal anomalies are computed by \eqref{eq:acdef} using the generator
\begin{equation}
R^{\rm UV}_{\mathcal{N}=1} = \tfrac{1}{3} R_{\mathcal{N}=2} +\tfrac{4}{3}I_{3}\ ,
\end{equation}
while for the IR conformal anomalies we use the generator
\begin{equation}
R^{\rm IR}_{\mathcal{N}=1} = \tfrac{1}{3} R_{\mathcal{N}=2} +\tfrac{4}{3}(I_{3}+ T_3)\ ,
\end{equation}
where we are assuming that the $\SU(2)_R$ and $\SU(2)_F$ generators are normalized in the same way, so that the respective structure constants are the same.  We now note that the following identities are true due the properties of the generators of $\SU(2)_R$ and $\SU(2)_F$
\begin{equation}
\text{Tr}(R_{\mathcal{N}=2}^2T_a) = \text{Tr}(T_3T_3T_3)= \text{Tr}(I_a)= \text{Tr}(T_a) = 0\ .
\end{equation}
With this at hand it is easy to show that
\begin{equation}\label{eq:TrRIRUV}
\text{Tr}[(R^{\rm IR}_{\mathcal{N}=1})^3) = \text{Tr}[(R^{\rm UV}_{\mathcal{N}=1})^3] - \tfrac{8}{9} k_F\,,\qquad \text{Tr}(R^{\rm IR}_{\mathcal{N}=1}) = \text{Tr}(R^{\rm UV}_{\mathcal{N}=1}) \ .
\end{equation}
Using these identities we arrive at the following simple relations between the UV and IR conformal anomalies
\begin{equation}\label{eq:aIRUV}
a_{\rm IR} = a_{\rm UV} - \tfrac{1}{4} k_F\ , \qquad\qquad c_{\rm IR} = c_{\rm UV} - \tfrac{1}{4} k_F \ .
\end{equation}
In unitary SCFTs one can show that $k_F>0$ so the result above is in harmony with the $a$-theorem.\footnote{Notice that there are stronger unitarity bounds on the flavor central charge given in Equations (4.16) and (4.17) of \cite{Beem:2013sza}.}

For theories with $a=c$, such as the large $N$ theories described by our holographic setup, the result \eqref{eq:aIRUV} can be written as
\be\label{c_ratio_SCFT}
\frac{c_{\rm IR}}{c_{\rm UV}} \,=\, 1-\frac{8}{9} \frac{k_F}{\text{Tr}[(R^{\rm UV}_{\mathcal{N}=1})^3]}  \,=\, 1+  \frac{\text{Tr}(R_{\mathcal{N}=2}T_3T_3)
}{\frac{1}{48}\text{Tr}(R_{\mathcal{N}=2}^3)+\text{Tr}(R_{\mathcal{N}=2}I_3I_3)}\ .
\ee

Now we can use the AdS/CFT dictionary to compare this expression with our supergravity results.
The relation between the SCFT symmetry generators and the supergravity vectors gauging that symmetry is
\begin{equation}\label{QFTgen_sugraA}
R_{\mathcal{N}=2}\,\to\, s A^0\,, \qquad I_{3} \,\to\, \tfrac{1}{g}A^3\,, \qquad T_3 \,\to\, \tfrac{\lambda}{g}A^8\ ,
\end{equation}
where the $1/g$ and $\lambda/g$ rescalings are introduced because in the conventions of Section \ref{sec:twoN=2vacua} the supergravity vectors $A^{1,2,3}$ and $A^{6,7,8}$ are gauging the $\SO(3)_1$ and $\SO(3)_c$ groups with gauge couplings $g$ and $g/\lambda$, respectively, while we have assumed that $I_a$ and $T_a$ have the same structure constants.
Moreover, $s$ is a real constant that is taking care of any potential rescaling of the $A^0$ gauge field in order to match CFT and supergravity conventions. It turns out that the specific value of this constant is not important for our analysis.

Using \eqref{QFTgen_sugraA}, the 't Hoof anomalies translate into coefficients of supergravity topological terms as
\be
\text{Tr}(R_{\mathcal{N}=2}^3) \to  s^3 d_{000}\ , \qquad \text{Tr}(R_{\mathcal{N}=2}I_3I_3) \to \tfrac{s}{g^2} d_{033}\ , \qquad \text{Tr}(R_{\mathcal{N}=2}T_3T_3)\to \tfrac{s\lambda^2}{g^2} d_{088}\ ,
\ee
where we are omitting a possible overall factor that will not play any role in our calculation. Therefore in supergravity language the expression in \eqref{c_ratio_SCFT} reads
\begin{equation}\label{prediction_sugra_formula}
\left(\frac{V_{\rm IR}}{V_{\rm UV}}\right)^{-3/2} =\, 1+ \lambda^2\frac{d_{088}}{\frac{s^2g^2}{48}d_{000}+ d_{033}}\ .
\end{equation}
In five-dimensional half-maximal supergravity, the coefficients $d_{000}$, $d_{033}$, $d_{088}$ are components of a symmetric tensor $d_{\mathcal{M}\mathcal{N}\mathcal{P}}$, with $\mathcal{M},\mathcal{N},\mathcal{P}=\{0,M\}=0,1,\ldots 5+\mathfrak{n}$, that controls the topological term. In particular, the gauge variation of the topological term contains $d_{\mathcal{M}\mathcal{N}\mathcal{P}} \mathcal{H}^\mathcal{M}\wedge \mathcal{H}^\mathcal{N}\wedge \delta A^{\mathcal{P}}$, where $\mathcal{H}^\mathcal{M}$ are covariant field strengths \cite{Schon:2006kz}. Crucially, the only non-zero components of the $d_{\mathcal{M}\mathcal{N}\mathcal{P}}$ tensor are $d_{0MN}=d_{M0N}=d_{MN0}=\eta_{MN}$.
Plugging $d_{000}=0$ and $d_{088}=-d_{033}$ into \eqref{prediction_sugra_formula} we obtain precisely the relation \eqref{ratioN=2case} we found in supergravity. Thus we find that the ratio of central charges of the UV and IR $\mathcal{N}=2$ SCFTs which we found in supergravity is precisely reproduced by the anomaly matching calculation above.

The discussion above also provides a field theory counterpart of the constant $\lambda$ entering in the supergravity embedding tensor and controlling the relation between the vevs of the operators triggering the flow.
Comparing \eqref{ratioN=2case} with \eqref{eq:aIRUV}, we obtain
\begin{equation}
\lambda^2 =  \frac{k_F}{4c_{\rm UV}}\,.
\end{equation}
The existence of the holographic RG flow imposes that $0<\lambda^2<1$ and it is important to understand whether this constraint can be understood from the dual large $N$ field theory. Unitarity of the SCFT immediately implies that $\lambda^2>0$, however we are not aware of any field theory argument for why one should find $\lambda^2<1$. It will be most interesting to understand better this condition and for which $\mathcal{N}=2$ SCFT it is obeyed.

%%%%%%%%%%%%%%%%%%%%%%%%%
\section{Holographic flows from  $\mathcal{N}=2$ to  $\mathcal{N}=1$ SCFTs}
\label{sec:Flow}
%%%%%%%%%%%%%%%%%%%%%%%%%

In this section we study holographic flows between an $\mathcal{N}=4$ AdS$_5$ vacuum and an $\mathcal{N}=2$ AdS$_5$ vacuum with a different cosmological constant. First we will provide the conditions for the existence of $\mathcal{N}=2$ AdS vacua, independently of whether there is also an $\mathcal{N}=4$ vacuum. Then we consider specific models allowing for an $\mathcal{N}=4$ AdS vacuum and study the existence of $\mathcal{N}=2$ AdS vacua. Finally, we construct domain wall solutions between such AdS vacua and discuss their holographic interpretation.

%%%%%%%%%%
\subsection{Conditions for $\mathcal{N}=2$ AdS$_5$ vacua}\label{sec:N=2AdSeqs}
%%%%%%%%%%

We start by providing general conditions for AdS$_5$ solutions of half-maximal gauged supergravity preserving eight supercharges, which have not been discussed in the literature so far. The only assumption we make is $\xi_M=0$.

The supersymmetries of an $\mathcal{N}=2$ AdS$_5$ solution transform as a doublet of $\SU(2)\simeq\USp(2)$, hence we need to identify the relevant $\USp(4)\to\SU(2)$ breaking of the R-symmetry of half-maximal supergravity. This was already discussed in~\cite{Cassani:2012wc} and we summarize it here.
The gravitino shift matrix \eqref{gravitino_shift} defines the SO(5) vector
\be\label{Xdefinition}
\widetilde{\vct}^{m} \,=\, \epsilon^{mnpqr} P_{np} P_{qr}\,,
\ee
with norm
\be
|\widetilde \vct| \,\equiv\, \sqrt{\widetilde{\vct}^{m}\widetilde{\vct}_{m}} \,=\, \sqrt{8\,(P^{mn}P_{mn})^2-16\,P^{mn}P_{np}P^{pq}P_{qm} }\;.
\ee
Let us focus on the generic case where this does not vanish (we will comment on the special case $\widetilde{\vct}=0$ at the end of this section).
 Then we can introduce a normalized vector
\be
\vct^m \,=\, \widetilde{\vct}^m\,/\,|\widetilde X|\ ,
\ee
which specifies an $\SO(4)$ subgroup of $\SO(5)$.
On the spinors, this defines a reduction USp(4) $\to {\rm SU}(2)_+\!\times {\rm SU}(2)_-$, where the plus and minus refer to the $\pm 1$ eigenvalues of $\vct_i{}^j = \vct_{m}\Gamma^{m}{}_i{}^j$. The supersymmetry preserved by our $\mathcal{N}=2$ AdS vacuum transforms under either one of these $\SU(2)$ factors. Without loss of generality we can choose $\SU(2)_+$, meaning that the supersymmetry parameters satisfy the projection
\be\label{USp4toUSp2proj}
\vct_i{}^j \epsilon_j = \epsilon_i\ .
\ee

Having identified the $\USp(4)\to \SU(2)$ breaking by means of the vector $X^m$, we find that the conditions for an $\mathcal{N}=2$ AdS$_5$ vacuum are:
\begin{align}
\hat{\xi}^{mn}X_n &= 0\ ,\label{condN=2AdS_1}\\[1mm]
\partial_\Sigma P^{mn} - \tfrac{1}{2}\varepsilon^{mnpqr}\partial_{\Sigma}P_{pq}X_r &= 0\ ,\label{condN=2AdS_2}\\[1mm]
\Sigma^3 \hat{\xi}^{ma} - \sqrt2 \hat{f}^{mna}X_n &= 0\ ,\label{condN=2AdS_3}\\[1mm]
(\delta^m_p - X^mX_p)(\delta^n_q - X^nX_q)\hat{f}^{pqa}  - \tfrac{1}{2}\varepsilon^{mn}{}_{pqr}\hat{f}^{pqa} X^r &= 0 \ .\qquad\qquad\label{condN=2AdS_4}
\end{align}
The proof is given below.
We observe that~\eqref{condN=2AdS_2} and \eqref{condN=2AdS_4} are self-duality conditions  on the four-dimensional space orthogonal to $X^m$.\footnote{We can derive some other, non-independent, relations. Contracting \eqref{condN=2AdS_2} with $X_n$ and using \eqref{condN=2AdS_1} we obtain $\hat{f}^{[mnp}X^{q]}= 0$, while contracting \eqref{condN=2AdS_3} with $X_m$ we find $\hat{\xi}^{an}X_n = 0$.}
The AdS radius is fixed by
\be\label{AdSradiusFromW}
\ell^{-1} = W \ ,
\ee
where
\begin{align}
W &= \sqrt{2\,P^{mn}P_{mn}- |\widetilde X|} \nn\\[1mm]
&= \sqrt{2\,P^{mn}P_{mn}- \sqrt{8\,(P^{mn}P_{mn})^2-16\,P^{mn}P_{np}P^{pq}P_{qm} }}
\ .\label{eq:ExprW}
\end{align}
As we will discuss in the next section, this expression for $W$ defines the superpotential driving  supersymmetric flows of the scalar fields. This is extremized at the AdS point.

It would be interesting to study the moduli space of the conditions above. This would most easily be done by exploiting the symmetry of the scalar manifold to set the undeformed vacuum at the origin and the unit vector $X^m$ to point in a chosen direction. However, this analysis goes beyond the scope of the present paper and we leave it for future work.

We can also discuss the spontaneous breaking of the gauge group in the $\mathcal{N}=2$ vacuum by looking at the scalar covariant derivative \eqref{ScalarCovDer_general}.
Working at leading order in the field fluctuations around the vacuum, separating the term along the vector $X^m$ from those transverse to it and using the supersymmetry conditions above, we get
\begin{align} \label{eq:Higgs}
X_m D\phi^{am} &= X_m \diff\phi^{am} + \tfrac{1}{\sqrt2}\Sigma^3\, \hat{\xi}^{am}\Pi_m^n \hat{A}_n + X_m\hat{f}^{mab} \hat{A}_b\ , \nn\\[1mm]
\Pi^m_n D\phi^{an} &= \Pi^m_n \diff \phi^{an} -  \Pi_n^m\hat f^{anp } \Pi_p^q \hat{A}_{q} + \Pi_n^m \hat f^{n a b } \hat{A}_{ b} - \hat\xi^{a m} \Big(A^0 + \tfrac{1}{\sqrt{2}} \Sigma^3 X_n \hat{A}^n \Big) \ ,
\end{align}
where $\Pi_m^n=\delta_m^n - X^nX_m$ projects on the subspace transverse to $X_m$.
The terms containing the $\hat{A}^{a}$ gauge vectors signal that all non-compact generators of the gauge group are spontaneously broken in the $\mathcal{N}=2$ vacuum and their gauge bosons acquire a mass via the St\"uckelberg mechanism. This is analogous to what happens in $\mathcal{N}=4$ AdS$_5$ vacua. The remaining terms give generically mass to some of the vectors of the form $\Pi^m_n A^n$ and to the combination $A^0 + \tfrac{1}{\sqrt{2}} \Sigma^3 X_m \hat{A}^m$. The \U(1) generated by the transformation
\be
A^0 \to  A^0 + \tfrac{1}{\sqrt{2}} \Sigma^3 \diff \lambda \ ,\qquad
A^m \to  A^m - X^m \diff \lambda \ ,
\ee
is unbroken and corresponds to the R-symmetry of the $\mathcal{N}=2$ vacuum. This also corresponds to the R-symmetry of the dual $\mathcal{N}=1$ SCFT.

%%%%%%%%%%%%%%%
\subsubsection*{Proof}
%%%%%%%%%%%%%%%

Let us derive the $\mathcal{N}=2$ supersymmetry conditions given above. Using the AdS conditions $\mathcal{A}'=\frac{1}{\ell}$ and $\Sigma'=\phi^x{}'=0$, the supersymmetry equations  \eqref{eq:susyeq1}--\eqref{eq:susyeq4} reduce to
\begin{align}
 \iu P_{i}{}^{j} \epsilon_j &= \tfrac{1}{\ell} \gamma_5 \epsilon_i \ , \label{eq:susyeqAdS1}\\[1mm]
\epsilon'_i  &= \tfrac{1}{2\ell} \epsilon_i \ , \label{eq:susyeqAdS2}\\[1mm]
\partial_\Sigma P^{ij} \epsilon_j &= 0 \ , \label{eq:susyeqAdS3}\\[1mm]
 P^{a\,ij} \epsilon_j & = 0
\label{eq:susyeqAdS4} \ .
\end{align}

Using \eqref{eq:susyeqAdS1} twice, we obtain
\be
- \widetilde\vct_i{}^j  \epsilon_j = \left[\tfrac{1}{\ell^2} -2 P_{mn} P^{mn} \right] \epsilon_i \ ,
\ee
and one can easily see that, as long as $\widetilde{\vct}$ does not vanish, and after making a harmless sign choice, this is equivalent to the $\USp(4)\to\SU(2)$ projection~\eqref{USp4toUSp2proj} together with eqs.~\eqref{AdSradiusFromW}, \eqref{eq:ExprW}~\cite{Cassani:2012wc}.

Eq.~\eqref{eq:susyeqAdS2} is trivially solved in terms of a constant spinor $\hat{\epsilon}_i$ as $\epsilon_i = \rme^{\frac{r}{2\ell}}\hat{\epsilon}_i$. However we must recall that \eqref{eq:susyeqAdS1}, \eqref{eq:susyeqAdS2} were derived from the gravitino variation assuming that the supersymmetry parameter $\epsilon_i$ does not depend on the $\mathbb{R}^{1,3}$ domain wall coordinates, therefore they only capture the Poincar\'e supersymmetry of AdS. When the conformal supersymmetries are also taken into account, one finds that the gravitino equation does not constrain the degrees of freedom in $\epsilon_i$ further than \eqref{USp4toUSp2proj}.
For this reason, the analysis from now on differs from the one in~\cite{Cassani:2012wc}, where only the Poincar\'e supersymmetries were considered.

The remaining two equations, namely \eqref{eq:susyeqAdS3} and \eqref{eq:susyeqAdS4}, constrain the embedding tensor and lead to the actual conditions for $\mathcal{N}=2$ vacua. Since they must hold on any spinor satisfying the projection~\eqref{USp4toUSp2proj}, we infer that
\begin{align}
\partial_\Sigma P^{ik} \left( \delta_k{}^j + \vct_k{}^j\right) &= 0 \ \nn,
\\[1mm]
 P^{a\,ik} \left( \delta_k{}^j + \vct_k{}^j\right) & = 0\ .
\end{align}
Recalling the definition of the shift matrices \eqref{gravitino_shift}, \eqref{gaugino_shift} and displaying the \SO(5) gamma matrices, these equations can be rewritten as
\begin{align}
\partial_\Sigma P^{mn} (\Gamma_{mn})^{ik} \left( \delta_k{}^j + \vct_p (\Gamma^p)_k{}^j\right) &= 0 \ , \label{eq:susyeqAdS3BIS}
\\[1mm]
\left(\sqrt{2}\,\Sigma^3\, \hat\xi^{am} (\Gamma_{m})^{ik} + \hat f^{amn} (\Gamma_{mn})^{ik}\right)
 \left( \delta_k{}^j + \vct_p (\Gamma^p)_k{}^j\right) & = 0
\label{eq:susyeqAdS4BIS}
\ .
\end{align}
Working out the contractions of the $\USp(4)$ indices, \eqref{eq:susyeqAdS3BIS} is equivalent to
\begin{align}
\partial_\Sigma P^{mn}X_n &= 0\ ,\nn\\
\partial_\Sigma P^{mn} - \tfrac{1}{2}\varepsilon^{mnpqr}\partial_{\Sigma}P_{pq}X_r &= 0\ .
\end{align}
The first can be combined with the identity $P^{mn}X_n=0$ (following from the definition of $X_n$ and the fact that $P^{m[n}P^{pq}P^{rs]}$ trivially vanishes in five dimensions) to give \eqref{condN=2AdS_1}, while the second is already the same as \eqref{condN=2AdS_2}.
Separating the different $\USp(4)$ representations, it is straightforward to see that \eqref{eq:susyeqAdS4BIS} is equivalent to \eqref{condN=2AdS_3}, \eqref{condN=2AdS_4}. This concludes our proof.

The derivation above assumed that $\widetilde{\vct}$ does not vanish. When $\widetilde{\vct}=0$  the solution may preserve eight Poincar\'e supercharges, which is the situation considered in Section~\ref{sec:onlyN=4vacua}. However, it may still be possible to have $\mathcal{N}=2$ AdS$_5$ vacua with vanishing $\widetilde{\vct}$. This still requires the existence of a unit vector $\vct$, however now unrelated to the $\widetilde{\vct}$ defined in \eqref{Xdefinition}, projecting out half of the spinor degrees of freedom as in \eqref{USp4toUSp2proj}. For this to be compatible with the gravitino equation we also need that $X_i{}^j$ and $P_i{}^j$ commute, which is equivalent to demanding $P^{mn}X_n=0$. The rest of the analysis of the supersymmetry equations is unchanged, hence conditions \eqref{condN=2AdS_1}--\eqref{condN=2AdS_4} still hold and the AdS radius is given by $\ell^{-1} = \sqrt{2P^{mn}P_{mn}}$.

%%%%%%%%%%%%%%
\subsection{Review of conditions for minimally supersymmetric flows}
%%%%%%%%%%%%%%

After having identified models admitting both $\mathcal{N}=4$ and $\mathcal{N}=2$ AdS$_5$ vacua, we will be interested in describing supersymmetric domain walls connecting them. Away from the fixed points, the domain wall should preserve just four Poincar\'e supercharges, namely the minimal amount of supersymmetry on $\mathbb{R}^{1,3}$.
The necessary and sufficient conditions for such domain walls in half-maximal supergravity were given in~\cite{Cassani:2012wc}.\footnote{The analysis of \cite{Cassani:2012wc} was restricted to an embedding tensor satisfying $\xi_M = 0$. Recall that we are also assuming this condition here as it is necessary for a fully supersymmetric AdS$_5$ vacuum.
Also note that in \cite{Cassani:2012wc} two superpotentials $W_\pm$ were constructed, depending on the preserved supersymmetry; without loss of generality here we choose $W=W_+$.} Here we summarize them.

The conditions use the same vector $X$ and the same superpotential $W$ defined in Section~\ref{sec:N=2AdSeqs}, however now the scalars are non-constant and depend on the radial coordinate. In addition to solving the flow equations
\begin{align}
	 \mathcal{A}'  &= \,W\,, \label{metricflow}\\[1mm]
	 \Sigma' &= -\Sigma^2\partial_\Sigma W\,, \label{sigmaflow}\\[1mm]
\phi^{x\,\prime} &= -3\,g^{xy}\,\partial_y W\,,	 \label{FloweqFromGaugino_MainText}
\end{align}
one has to impose the following constraints along the flow:
\begin{align}
\partial_\Sigma \vct^m &= 0\ ,\label{condition_dSigmaX}\\[1mm]
\partial_\Sigma\left(W^{-1}P_+^{mn}\right) &=0\ , \label{condition_Pplus}\\[1mm]
\hat\xi^{am}\vct_{m} &= 0\ ,\label{condition_xi}\\[1mm]
\hat{f}_+^{mna} - \tfrac{4}{W^2} P_{+pq} \hat{f}_+^{pqa}\, P_+^{mn}  & = 0\ ,\label{condition_f}
\end{align}
where we have introduced
\be
P_+^{mn} = \tfrac{1}{2} \left(P^{mn}- \tfrac{1}{2}\epsilon^{mnpqr}P_{pq}X_r \right)\
\ee
and
\be
\hat{f}_+^{mna} = (\delta^m_p - X^mX_p)(\delta^n_q - X^nX_q)\hat{f}^{pqa}  - \tfrac{1}{2}\varepsilon^{mn}{}_{pqr}\hat{f}^{pqa} X^r\ ,
\ee
both living in the four-dimensional space orthogonal to $X$ and being anti-self-dual.\footnote{The ``$+$'' subscript comes from the original definitions in~\cite{Cassani:2012wc}. Although expressed in a slightly different form, these constraints are equivalent to eqs.~(3.29), (3.31), (3.32) in \cite{Cassani:2012wc}.}

The superpotential \eqref{eq:ExprW} can also be written as $W = 2\sqrt{P_+^{mn}P_{+mn}}$. One can then use \eqref{condition_Pplus} to show that $\partial_\Sigma P_+$ is proportional to $P_+$, and is therefore analogous to \eqref{condition_f}.

We are interested in constructing domain wall solutions interpolating between an $\mathcal{N}=4$ and an $\mathcal{N}=2$ AdS$_5$ vacuum. Thus one of the fixed points has to satisfy the restrictive $\mathcal{N}=4$ conditions \eqref{eq:N=4condA}--\eqref{eq:N=4condD}.\footnote{Notice that the vector $\widetilde X^m$ has to vanish there, so that the four Poincar\'e supersymmetries preserved along the flow can be enhanced to eight (plus the conformal supersymmetries).}
The other fixed point instead has to satisfy the $\mathcal{N}=2$ conditions \eqref{condN=2AdS_1}--\eqref{condN=2AdS_4}. One can see that the latter are in fact equivalent to the constraints \eqref{condition_dSigmaX}--\eqref{condition_f}, together with the condition that the superpotential is extremized.We now proceed to discuss two explicit examples which display all these features.

%%%%%%%%%%%
\subsection{A model with one $\mathcal{N}=2$ vacuum}
\label{sec:oneN=2vacuum}
%%%%%%%%%%%

An example of a supersymmetric domain wall solution connecting a maximally supersymmetric AdS$_5$ vacuum to an $\mathcal{N}=2$ AdS$_5$ vacuum is the well-known Freedman-Gubser-Pilch-Warner (FGPW) flow \cite{Freedman:1999gp}. This was originally constructed in the $\SO(6)$ maximal supergravity, where the UV vacuum is the standard $\SO(6)$ invariant critical point, while the IR $\mathcal{N}=2$ vacuum
 is the one first found in \cite{Khavaev:1998fb}. As discussed in  \cite{Freedman:1999gp} this domain wall solution can also be described in half-maximal gauged supergravity by a model with two $\mathcal{N}=4$ vector multiplets and a gauging determined by the truncation of $\SO(6)$ maximal supergravity. Here we extend the FGPW model allowing for a more general gauging.
We could also allow for an arbitrary number of vector multiplets as done in Section~\ref{sec:onlyN=4vacua} when studying flows between two $\mathcal{N}=4$ vacua (see~\cite{Corrado:2002wx} for such an extension of the FGPW model), however all essential features of the flow are already captured by a model with two multiplets, so we restrict to that.

We choose the embedding tensor as
\be\label{embtensorN=2vacuum}
f^{123} = g\ ,\qquad \xi^{45} = - \tfrac{g}{\sqrt{2}}\ , \qquad \xi^{67} = - \sqrt{2}\,g \rho^{-1}\ ,
\ee
where $g$ and $\rho$ are parameters.
The vectors $A^1,A^2,A^3$ gauge $\SU(2)$, $A^0$ gauges $\U(1)$, while $A^4,A^5,A^6,A^7$ are eaten up by tensor fields.

The FGPW model obtained by truncating $\SO(6)$ maximal supergravity has $\rho=2$, so that $\xi^{67}=\xi^{45}$. In this case the fully superymmetric vacuum has a complex modulus, parameterizing the space $\SU(1,1)/\U(1)$.
Since the conditions of Section~\ref{sec:N=4AdSconditions} are satisfied, we have a fully supersymmetric solution at the origin of the scalar manifold for any value of $\rho$. In order to obtain an $\mathcal{N}=2$ vacuum at some other point of the scalar manifold, we break the $\SO(3)$ rotations in the $1,2,3$ directions by mixing the $1,2$ and $6,7$ directions on the scalar manifold.
We thus parameterize the  $ \frac{\SO(5,2)}{\SO(5)\times \SO(2)}$ coset representative as\footnote{We could introduce two different scalars but the $\mathcal{N}=2$ vacuum conditions would set them equal.}
\begin{equation}\label{eq:N=2ansatz}
{\cal{V}} \,=\,\rme^{-2\phi\, t_{16}-2\phi\, t_{27}}= \left(\begin{matrix}
\cosh\phi & 0 & 0 & 0 & 0 & -\sinh\phi & 0 \\
 0 & \cosh\phi & 0 & 0 & 0 & 0 & -\sinh\phi  \\
0 & 0 & 1 & 0 & 0 & 0 & 0 \\
0 & 0 & 0 & 1 & 0 & 0 & 0 \\
 0 & 0 & 0 & 0 & 1 & 0 & 0 \\
 -\sinh\phi & 0 & 0 & 0 & 0 & \cosh\phi & 0 \\
0 & -\sinh\phi & 0 & 0 & 0 & 0 & \cosh\phi \\
 \end{matrix}\right)\ .
\end{equation}
The dressed embedding tensor \eqref{dressed_emb_tensor} then reads:
\begin{equation} \label{eq:flowingcouplings}\begin{array}{lll}
\hat\xi^{12} =  -\sqrt2 g \rho^{-1} \sinh^2 \phi \ ,& & \qquad \hat f^{123} =  g \cosh^2 \phi \ , \\[2mm]
\hat\xi^{45} =  - \tfrac{g}{\sqrt2} \ ,& &\qquad  \hat f^{137} = -\hat f^{236} = g\sinh\phi\cosh\phi \ , \\[2mm]
\hat\xi^{17} =  -\hat\xi^{26} = \sqrt2  g \rho^{-1}\sinh \phi \cosh \phi \ ,& &\qquad  \hat f^{367} = g \sinh^2\phi \ ,\\[2mm]
\hat\xi^{67} =  -\sqrt2 g \rho^{-1}\cosh^2\phi \ ,  & &
\end{array} \end{equation}
where by 6,7 we are denoting the values taken by the $a$ index. For the unit vector defining the $\USp(4)\to \SU(2)$ projection of the supersymmetries we find $X^m = {\rm sign}(\rho) \delta^m_3$.

The metric on the space spanned by the scalars $\Sigma,\phi$ in this case is
\be
\diff s^2 \, =\, 3\,\Sigma^{-2}\diff \Sigma^2 + 2\,\diff\phi^2 \ ,
\ee
while the scalar potential is
\be\label{scal_pot_N=4N=2_vacua}
V \, =\, g^2 \cosh^2\phi \left[\Sigma^{4}\rho^{-2}\sinh^2\phi -\Sigma + \tfrac{1}{4}\Sigma^{-2}\left(\cosh(2\phi)-3\right)\right]    \ .
\ee

The $\mathcal{N} =2$ vacuum conditions \eqref{condN=2AdS_1}, \eqref{condN=2AdS_4} are satisfied automatically. Eq.~\eqref{condN=2AdS_2} gives
\be
\cosh^2\phi = \Sigma^3\left(1- 2|\rho|^{-1}\sinh^2\phi \right)\ ,
\ee
while \eqref{condN=2AdS_3} yields
\be
(1-|\rho|^{-1}\Sigma^3)\sinh\phi = 0 \ .
\ee
In addition to the $\mathcal{N}=4$ AdS$_5$ solution
\be
\Sigma = 1\ , \qquad \phi = 0\ , \qquad V = -\tfrac{3}{2}g^2\ ,
\ee
we obtain the $\mathcal{N}=2$ AdS$_5$ solution
\be\label{eq:IRN=2AdS5}
\Sigma^3 = |\rho| \ , \qquad \rme^{2\phi} =  \tfrac{1}{3}\left(1+2|\rho| \pm 2\sqrt{\rho^2+|\rho|-2}\right)\, ,\qquad V = -\tfrac{1}{6}g^2|\rho|^{-2/3}(2+|\rho|)^2\ .
\ee
Note that the latter only exists for $|\rho| > 1$ since only then we have a real scalar $\phi$. For $|\rho|\to 1$ the $\mathcal{N}=2$ AdS$_5$ vacuum merges with the $\mathcal{N}=4$ vacuum at the origin.

In the $\mathcal{N}=2$ vacuum most of the gauge symmetries are broken.
From \eqref{eq:Higgs} we see that the non-trivial scalar covariant derivatives around this vacuum are:
\begin{align}
D\phi^{63} &= \diff \phi^{63} \pm g \sqrt{\tfrac{|\rho|-1}{3}} A^2\ , \nn\\[1mm]
D\phi^{73} &= \diff \phi^{73} \mp g \sqrt{\tfrac{|\rho|-1}{3}} A^1\ , \nn\\
D(\phi^{62}-\phi^{71}) &= \diff(\phi^{62}-\phi^{71}) \mp \tfrac{2}{3}g \sqrt{\rho^2+|\rho|-2} \left(\sqrt2\rho^{-1}A^0 + A^3 \right) \ .
\end{align}
%
%while the combination $(\phi_{6 2}+\phi_{71})$ remains uncharged. Also the scalar $\phi_{61} = \phi_{72}=\phi$ that is involved in the flow is uncharged.
The vector fields on the right hand side get a mass through the St\"uckelberg mechanism.
The vectors in the first two lines are just two of the gauge vectors of the gauged ${\rm SU}(2)$.
The $\mathcal{N}=2$ vacuum is invariant under the combination $\left( \frac{1}{\sqrt2 }\rho^{-1} A^0 - A^3 \right)$, corresponding to the $\U(1)$ R-symmetry.

We can now move on to study the supersymmetric flow connecting the $\mathcal{N}=4$ and the $\mathcal{N}=2$ vacua. The superpotential reads
\be
W \, =\,  \tfrac{g}{3} \, \Sigma^{-1}\cosh^2\phi +  \tfrac{g}{6}\, \Sigma^2 \left( 1-2|\rho|^{-1}\sinh^2\phi \right)  \ ,
\ee
where we are assuming $g> 0$. It is easy to check that with the parameterization \eqref{eq:N=2ansatz} of the coset representative, the constraints \eqref{condition_dSigmaX}--\eqref{condition_f} are satisfied. This means that it is consistent to assume that the only flowing scalars are $\Sigma,\phi$. One can also check that the scalar potential and the superpotential are related as
\begin{equation}
V = \tfrac{3}{2}\,\Sigma^2 (\partial_{\Sigma}W)^2+\tfrac{9}{4} (\partial_{\phi}W)^2- 6 W^2\ .
\end{equation}
in agreement with \eqref{VfromW_general}. The flow equations \eqref{flow_sigma}, \eqref{flow_phi} for the scalar fields read
\begin{align}\label{floweqsN=4toN=2}
\Sigma' \,& =\, \tfrac{g}{3}\left[ \cosh^2\phi + \Sigma^3\left(2|\rho|^{-1}\sinh^2\phi -1 \right) \right]  \ , \\[1mm]
\phi' \,& =\, \tfrac{g}{2} \left( |\rho|^{-1}\Sigma^2 - \Sigma^{-1} \right)\sinh(2\phi) \ .
\end{align}
From now on we assume without loss of generality that $\rho>0$ so that we can remove the absolute values.

Let us call the operators dual to the two scalars $\mathcal{O}_{\phi}$ and $\mathcal{O}_{\Sigma}$. Expanding around the $\mathcal{N}=4$ vacuum one finds that the dimensionless masses of the two scalars are
\begin{equation}\label{eq:massesphiSigmaN2}
m^2_{\phi}\ell^2 = -4(1-\rho^{-2})\,, \qquad m^2_{\Sigma}\ell^2 = -4 \;.
\end{equation}
Using the standard AdS/CFT relation $m^2\ell^2 = \Delta(\Delta-4)$ this implies that the dimensions of the dual operators are\footnote{One could in principle choose the other root of the quadratic equation for $\Delta_{\mathcal{O}_{\phi}}$, i.e. $\Delta_{\mathcal{O}_\phi}= 2-\frac{2}{\rho}$. This however violates the unitarity bound, $\Delta>1$, for $1<\rho<2$. Moreover for $\rho=2$ we know from the FGPW model that $\Delta_{\mathcal{O}_{\phi}}=3$ which is obeyed for the choice in \eqref{UVAdS}.}
\begin{equation}\label{UVAdS}
\Delta_{\mathcal{O}_{\phi}} = 2+\frac{2}{\rho}\;, \qquad\qquad \Delta_{\mathcal{O}_{\Sigma}} = 2 \;.
\end{equation}
Along the RG flow there is operator mixing and in the IR SCFT we have two new eigenstates of the dilatation operator. The corresponding operator dimensions are
\begin{equation}\label{}
\Delta_{\mathcal{O}_{1}} = 3+\sqrt{25-\frac{72}{2+\rho}}\;, \qquad\qquad \Delta_{\mathcal{O}_{2}} = 1+\sqrt{25-\frac{72}{2+\rho}} \;.
\end{equation}
For any $\rho>1$ we have that $\Delta_{\mathcal{O}_{1}} >4$ and thus this is always an irrelevant operator. For $\Delta_{\mathcal{O}_{2}} $ one finds
\begin{equation}\label{}
\begin{split}
2\leq \Delta_{\mathcal{O}_{2}} \leq 4, \qquad 1 \leq \rho \leq 2.5\;, \\
4<\Delta_{\mathcal{O}_{2}} \leq6, \qquad 2.5 < \rho < \infty\;.
\end{split}
\end{equation}

The ratio of central charges (in the planar limit) of the dual SCFTs is
\begin{equation}\label{cratiosugra}
\frac{c_{{\rm IR}}}{c_{{\rm UV}}} = \left(\frac{V_{\mathcal{N}=2}}{V_{\mathcal{N}=4}}\right)^{-3/2}  =  \frac{27\rho}{(2+\rho)^3}\;.
\end{equation}
Since the $\mathcal{N}=2$ vacuum only exists for $\rho>1$, we find that the well-known $27/32$ ratio of central charges \cite{Freedman:1999gp,Tachikawa:2009tt} is realized only for $\rho=2$. As already noticed, this value is also exactly the one where one finds a modulus for the $\mathcal{N}=4$ vacuum, corresponding to a marginal coupling in the dual SCFT.

Let us compare the ratio in \eqref{cratiosugra} with the ratio of central charges from equation (2.22) in \cite{Bah:2012dg} where we fix $z=1$ for the UV theory (this corresponds to the Maldacena-N\'u\~nez $\mathcal{N}=2$ solution) and the goal is to map the parameter $z$ from \cite{Bah:2012dg} to the parameter $\rho$ in \eqref{cratiosugra}. From \cite{Bah:2012dg} we find
\begin{equation}\label{cratioCFT}
\frac{c_{{\rm IR}}}{c_{{\rm UV}}} =\frac{9z^2-1+(1+3z^2)^{3/2}}{16z^2}\;.
\end{equation}
One can now find a map between $z^2$ and $\rho$. The explicit expression is not very illuminating but one finds that $z^2=0$ is mapped to $\rho=2$ and $z^2=1$ is mapped to $\rho=1$. Moreover the map is monotonic, i.e. if we restrict ourselves to $0\leq z^2 \leq 1$ we have to restrict $\rho$ to be in the range $2\geq \rho \geq 1$. This suggest that our model with two vector multiplets may describe holographic RG flows between the $\mathcal{N}=2$ Maldacena-N\'u\~nez vacuum and some of the $\mathcal{N}=1$ vacua with $|z|<1$ studied in \cite{Bah:2012dg}.

\begin{figure}[h]
\centering
\includegraphics[width=7.4cm]{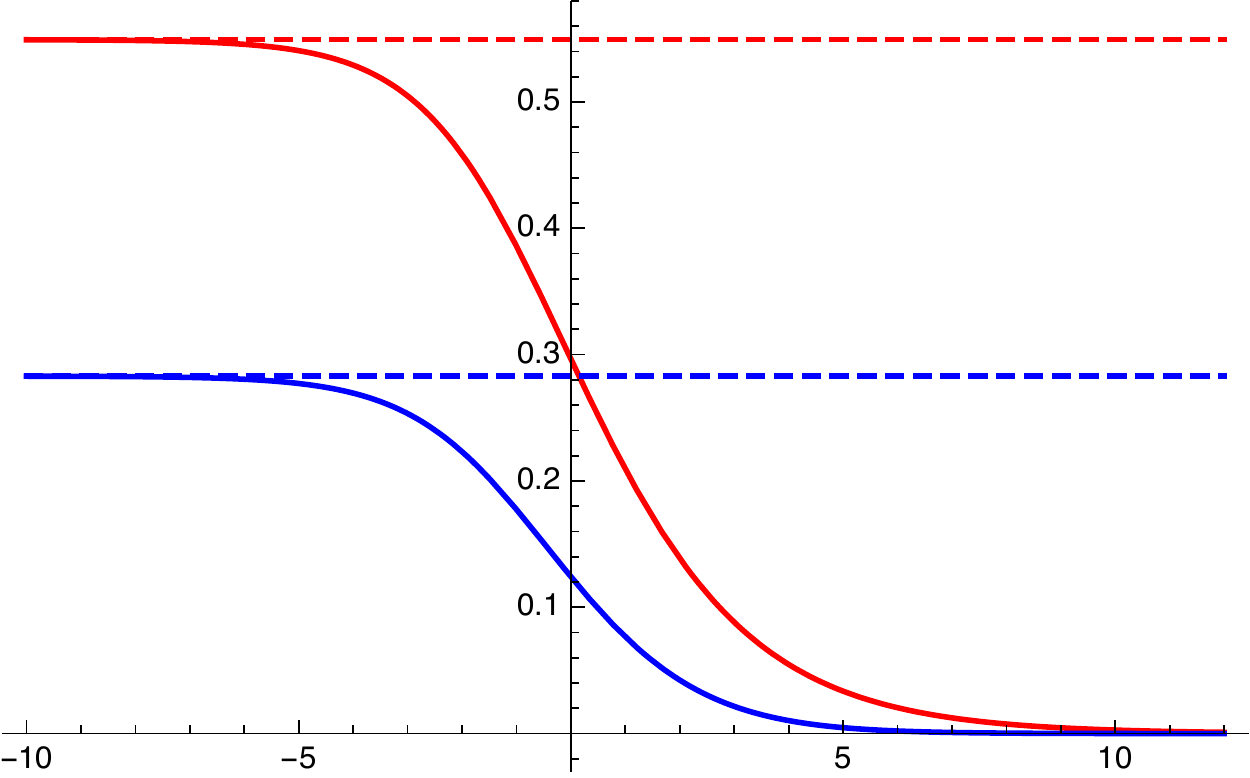} \qquad\qquad \includegraphics[width=7cm]{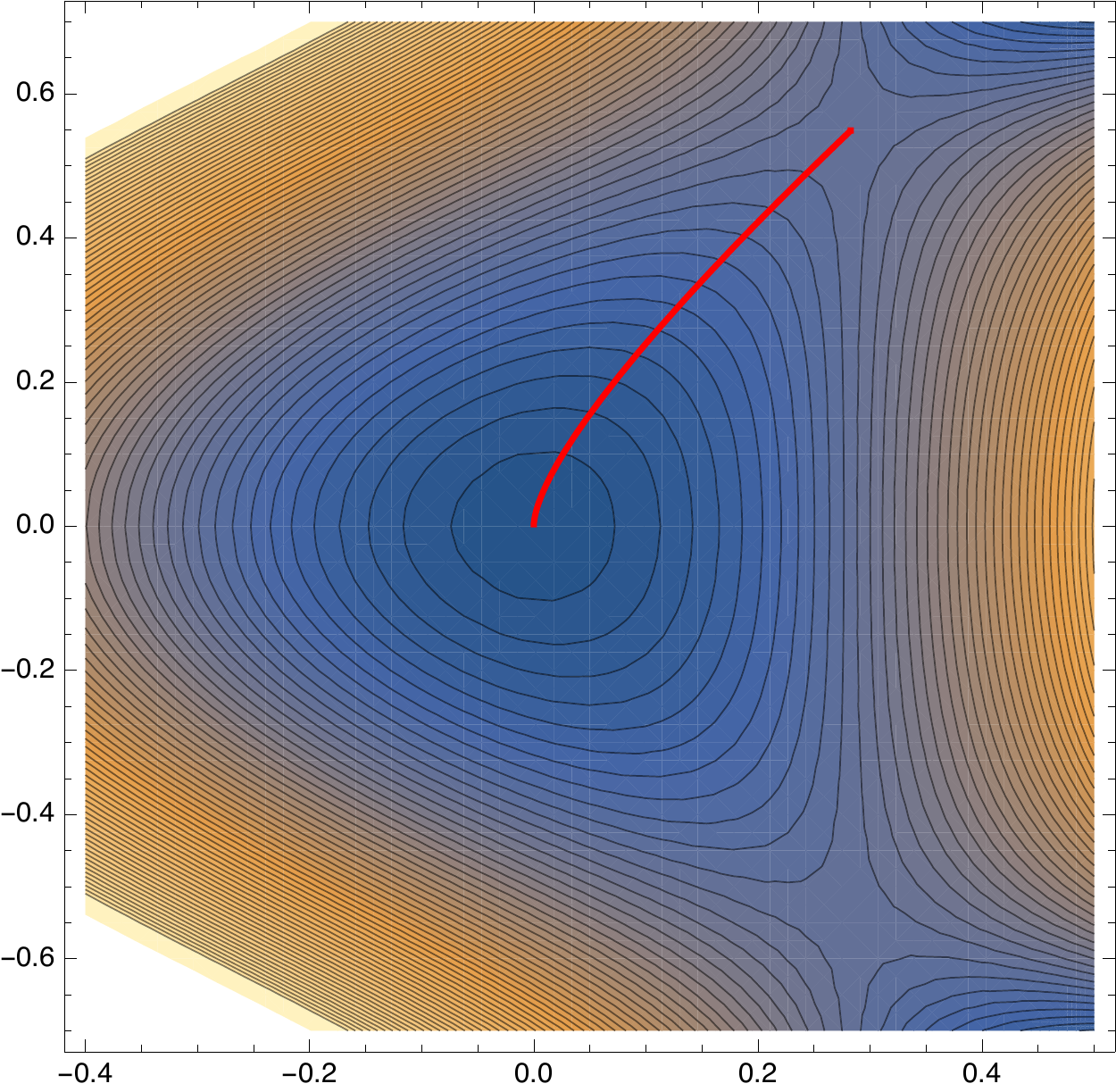}
\caption{\label{fig:2732} \textit{Left}: Numerical solutions for $\phi(r)$ (red solid line) and $\sqrt6\log\Sigma(r)$ (blue solid line). The dashed red/blue lines are the values for the scalars at the IR $\mathcal{N}=2$ AdS$_5$ vacuum at $r\to -\infty$. The UV $\mathcal{N}=2$ AdS$_5$ vacuum is at $r\to \infty$ with $\phi=\sqrt6\log\Sigma=0$ there. We have fixed $g=1$ and $\rho=2$. \textit{Right}: A contour plot of the superpotential as a function of the scalars $\sqrt6\log\Sigma$ (horizontal axis) and $\phi$ (vertical axis) together with a parametric plot of the numerical solution for the scalars from the left panel.}
\end{figure}

The flow equations \eqref{floweqsN=4toN=2} for this model can be integrated numerically. This is illustrated in Figure~\ref{fig:2732}. It is clear from this figure that there is a smooth domain wall solution which interpolates between the $\mathcal{N}=4$ and $\mathcal{N}=2$ AdS$_5$ vacua.

To understand better what drives the flow we can expand the BPS flow equations near the $\mathcal{N}=4$ AdS$_5$ vacuum in the UV. The linearized expansion of the BPS equations depends on the value of $\rho$. For  $\rho>2$ we find
\begin{equation}\label{eq:FGphisigmag2}
\phi \approx c_\phi\,\rme^{-(2-2/\rho)r/\ell}\,, \qquad\qquad \Sigma \approx 1 + c_\Sigma\, \rme^{-2r/\ell}\,,
\end{equation}
while for $1<\rho<2$ the result is
\begin{equation}\label{eq:FGphisigmal2}
\phi \approx c_\phi\,\rme^{-(2-2/\rho)r/\ell}\,, \qquad\qquad \Sigma \approx 1 +c_\phi^2\frac{2+\rho}{3(2-\rho)}\rme^{-4r(1-\rho^{-1})/\ell} + c_\Sigma\, \rme^{-2r/\ell}\,.
\end{equation}
Using \eqref{UVAdS} we conclude that the RG flow is driven by a source term for the operator $\mathcal{O}_\phi$ proportional to the constant $c_\phi$. The constant $c_{\Sigma}$ is related to the vev of the operator $\mathcal{O}_\Sigma$ which is dynamically generated along the RG flow. The expression in \eqref{eq:FGphisigmag2} has the expected form for scalar fields with masses as in \eqref{eq:massesphiSigmaN2}. The result in \eqref{eq:FGphisigmal2} however is different since for $1<\rho<2$ one should keep quadratic (and higher order) terms in $\phi$ in the linearized expansion of the differential equation for the scalar $\Sigma$ in \eqref{floweqsN=4toN=2}. 

The case $\rho=2$ should be treated separately and the linearized expansion of the BPS equations near the $\mathcal{N}=4$ AdS$_5$ then reads
\begin{equation}
\phi \approx c_\phi\,\rme^{-r/\ell}\,, \qquad\qquad \Sigma \approx 1 + \tfrac{4}{3\ell}c_\phi^2\, r \rme^{-2r/\ell} + c_\Sigma\, \rme^{-2r/\ell}\,.
\end{equation}
This is the behavior of an RG flow triggered by sources for operators of dimensions $3$ and $2$. This behavior was also observed in Section 5 of \cite{Freedman:1999gp}. The regular numerical solution displayed in Figure~\ref{fig:2732} fixes a particular relation between the constants $c_{\phi}$ and $c_{\Sigma}$ which depends on the value of $\rho$.

Now we turn our attention to reproducing the ratio \eqref{cratiosugra} between the central charges from a field theory argument. This can be viewed as a generalization of the results in \cite{Tachikawa:2009tt} which is reproduced by selecting $\rho=2$ above. To this end suppose that we have a deformation of the $\mathcal{N}=2$ SCFT dual to the $\mathcal{N}=4$ AdS$_5$ vacuum in the UV which is such that the resulting RG flow ends in an $\mathcal{N}=1$ SCFT with a superconformal R-symmetry given by the following linear combination of the Cartan generators of the UV $\SU(2)\times \U(1)$ R-symmetry
\be\label{eq:RIRalphaQFT}
R^{\rm IR}_{\mathcal{N}=1} = \frac{1+\alpha}{3}R_{\mathcal{N}=2} +\frac{4-2\alpha}{3} I_3\,.
\ee
Using this superconformal R-symmetry and the anomaly relations in \eqref{eq:acdef} one readily finds the following relation between the UV and IR central charges
\begin{equation}\label{eq:acIRalpha}
\begin{split}
a_{\rm IR} &= (1+\alpha^3) a_{\rm UV}-\tfrac{3}{4}\alpha^2(\alpha+1)c_{\rm UV}\,, \\
c_{\rm IR} &= \alpha(\alpha^2-1) a_{\rm UV}+\tfrac{1}{4}(\alpha+1)(4-3\alpha^2)c_{\rm UV}\,.
\end{split}
\end{equation}
For $\alpha=1/2$ the result above reproduces the anomaly calculation in \cite{Tachikawa:2009tt}. When the UV theory has $a_{\rm UV}=c_{\rm UV}$, such as in SCFTs with a weakly coupled gravity dual, one finds that the relations in \eqref{eq:acIRalpha} reduce to
\begin{equation}
a_{\rm IR} = c_{\rm IR} = \tfrac{1}{4}(\alpha+1)(\alpha-2)^2\, a_{\rm UV}\,.
\end{equation}
This suggests that to reproduce the supergravity result found in \eqref{cratiosugra} above we have to make the identification\footnote{Unitarity and the $a$-theorem imply that $2>\alpha>0$ which is mapped to the range $\infty>\rho>1$ in the supergravity analysis.}
\begin{equation}\label{eq:rhoalpharelation}
\rho = \frac{2+2\alpha}{2-\alpha}\,.
\end{equation}
This indeed turns out to be the case since as we show in Appendix~\ref{app:genTW} the combination of gauge fields which are massless at the IR $\mathcal{N}=2$ vacuum in \eqref{eq:IRN=2AdS5} corresponds to the generator
\begin{equation}\label{eq:RIRsugra}
R^{\rm IR}_{\mathcal{N}=1} = \frac{\rho}{\rho+2}R_{\mathcal{N}=2} +\frac{4}{\rho+2} I_3\,,
\end{equation}
which after the identification in \eqref{eq:rhoalpharelation} reduces to \eqref{eq:RIRalphaQFT}. As an additional consistency check one can show that the charge of the scalar $\phi$ under the supergravity gauge field corresponding to the UV superconformal R-symmetry generator, i.e.\ the one in \eqref{eq:RIRalphaQFT} with $\alpha=0$, is $\frac{4}{3}(1+\rho^{-1})$. This should correspond to the superconformal R-charge of the operator $\mathcal{O}_{\phi}$ in the dual SCFT. It is generally expected that operators dual to supergravity scalar fields belong to chiral multiplets and thus the conformal dimension of $\mathcal{O}_{\phi}$ should be determined by its R-charge via the relation
\begin{equation}
\Delta_{\phi} = \frac{3}{2}\times \frac{4}{3}(1+\rho^{-1}) = 2+\frac{2}{\rho}\,.
\end{equation}
It is reassuring to find that this result nicely agrees with the one obtained in \eqref{UVAdS} from an explicit evaluation of the mass of the scalar $\phi$.

%%%%%%%%%%%%%%
\subsection{A model with two $\mathcal{N}=2$ vacua}
\label{subsec:2N=2vacua}
%%%%%%%%%%%%%%

We now consider a more involved model displaying an $\mathcal{N}=4$ vacuum and two distinct $\mathcal{N}=2$ vacua. Since this is similar to the previous example we studied we keep the presentation short. We take four vector multiplets and choose the embedding tensor as
\be
f^{123} = g\ ,\qquad \xi^{45} = - \tfrac{g}{\sqrt{2}}\ , \qquad \xi^{67} = - \sqrt{2}\,g \rho_1^{-1}\ , \qquad \xi^{89} = - \sqrt{2}\,g \rho_2^{-1}\ .
\ee
For simplicity, we assume $g>0$, $\rho_1>0$, $\rho_2>0$.
We parameterize an $ \frac{\SO(5,4)}{\SO(5)\times \SO(4)}$ coset element in terms of the scalar fields $\phi$, $\chi$ as
\begin{align}
\mathcal{V}&= \ \rme^{-2\phi\cos\chi \,(t_{16}+ t_{27}) -2\phi\sin\chi\,(t_{18}+ t_{29})}\nn\\[2mm]
&\!=\!{\scriptsize\left(\begin{matrix}
\ch\phi & 0 & 0 & 0 & 0 & -\sh\phi\cos\chi & 0 & -\sh\phi\sin\chi & 0\\
 0 & \ch\phi & 0 & 0 & 0 & 0 & -\sh\phi\cos\chi  & 0 & -\sh\phi\sin\chi\\
0 & 0 & 1 & 0 & 0 & 0 & 0 & 0 & 0\\
0 & 0 & 0 & 1 & 0 & 0 & 0& 0 & 0 \\
 0 & 0 & 0 & 0 & 1 & 0 & 0& 0 & 0 \\
\! -\sh\phi \cos\chi & 0 & 0 & 0 & 0 & \!\ch\phi\cos^2\chi + \sin^2\chi & 0& \sh^2\frac{\phi}{2}\sin2\chi & 0 \\
0 &\! -\sh\phi\cos\chi & 0 & 0 & 0 & 0 & \!\ch\phi\cos^2\chi + \sin^2\chi & 0 & \sh^2\frac{\phi}{2}\sin2\chi\\
\! -\sh\phi \sin\chi & 0 & 0 & 0 & 0 & \sh^2\frac{\phi}{2}\sin2\chi & 0& \!\ch\phi\sin^2\chi + \cos^2\chi & 0 \\
0 &\! -\sh\phi\sin\chi & 0 & 0 & 0 & 0 & \sh^2\frac{\phi}{2}\sin2\chi & 0 & \!\ch\phi\sin^2\chi + \cos^2\chi
 \end{matrix}\right)}\nn
\end{align}

Again we have $X^m = \delta^m_3$. In addition to the usual $\mathcal{N}=4$ vacuum at the origin with cosmological constant $V=-\frac{3}{2}g^2$, we obtain two $\mathcal{N}=2$ vacua by solving the supersymmetry conditions in a way similar to the example in Section~\ref{sec:oneN=2vacuum}. The first $\mathcal{N}=2$ vacuum is
\be\label{eq:N2vacuum1}
\Sigma^3 =\rho_1\ ,\quad \chi = 0\ ,\quad \rme^{2\phi} =  \tfrac{1}{3}\Big(1+2\rho_1 + 2\sqrt{\rho_1^2+\rho_1-2}\,\Big)\ , \quad V_{\mathcal{N}=2}^{(1)} = -\tfrac{g^2}{6}\rho_1^{-2/3}(2+\rho_1)^2\ ,\
\ee
while the second is
\be\label{eq:N2vacuum2}
\Sigma^3 =\rho_2\ ,\quad \chi = \tfrac{\pi}{2}\ ,\quad \rme^{2\phi} = \tfrac{1}{3}\Big(1+2\rho_2 + 2\sqrt{\rho_2^2+\rho_2-2}\,\Big)\ ,  \quad V_{\mathcal{N}=2}^{(2)} = -\tfrac{g^2}{6}\rho_2^{-2/3}(2+\rho_2)^2\ .\
\ee
Note that for the two vacua to be distinct we need $\rho_1\neq\rho_2$. In addition it is simple to find the ratio of central charges of the dual SCFTs. If we assume that $\rho_2>\rho_1>1$ we find that the $\mathcal{N}=4$ vacuum is in the UV, the vacuum in \eqref{eq:N2vacuum1} is intermediate and the vacuum in \eqref{eq:N2vacuum2} is in the deep IR,
\begin{equation}\label{cratiosugraN2}
\frac{c_{{\rm IR}}^{(1)}}{c_{{\rm UV}}} = \left(\frac{V_{\mathcal{N}=2}^{(1)}}{V_{\mathcal{N}=4}}\right)^{-3/2}  =  \frac{27\rho_1}{(2+\rho_1)^3}\,, \qquad\qquad \frac{c_{{\rm IR}}^{(2)}}{c_{{\rm UV}}} = \left(\frac{V_{\mathcal{N}=2}^{(2)}}{V_{\mathcal{N}=4}}\right)^{-3/2}  =  \frac{27\rho_2}{(2+\rho_2)^3}\;.
\end{equation}

The metric on the space spanned by the three scalars $\Sigma,\phi,\chi$ is
\be
\diff s^2 \, =\, 3\,\Sigma^{-2}\diff \Sigma^2 + 2\,\diff\phi^2  + 2\sinh^2\phi\,\diff \chi^2\ .
\ee

The expression for the scalar potential is not particularly illuminating, however it can easily be recovered using \eqref{VfromW_general} and the superpotential
\be
W \,=\, \tfrac{g}{3} \, \Sigma^{-1}\cosh^2\phi + \tfrac{g}{6}\, \Sigma^2 \left[\, 1 - 2\left( \rho_1^{-1} \cos^2\chi + \rho_2^{-1}\sin^2\chi \right)\sinh^2\phi\, \right]\ .
\ee

Let us now discuss possible supersymmetric flows connecting the three supersymmetric vacua in this model.
The constraints \eqref{condition_dSigmaX}--\eqref{condition_f} are satisfied, so a flow involving $\Sigma,\phi,\chi$ will not require switching on other scalars.
The superpotential above generates the following flow equations for the scalar fields:
\begin{align}
\Sigma' \,& =\, \tfrac{g}{3}\left[ \,\cosh^2\phi + 2\Sigma^3\left( \rho_1^{-1}\cos^2\chi +\rho_2^{-1}\sin^2\chi \right)\sinh^2\phi -\Sigma^3\,  \right]  \ , \nn\\[1mm]
\phi' \,& =\, \tfrac{g}{2} \left[ \,\Sigma^2 \left( \rho_1^{-1}\cos^2\chi +\rho_2^{-1}\sin^2\chi \right) - \Sigma^{-1} \,\right]\sinh(2\phi)  \ , \nn\\[1mm]
\chi' \,& =\, \tfrac{g}{2} \left( \,\rho_2^{-1} - \rho_1^{-1}\, \right) \Sigma^2 \sin(2\chi)  \ .
\end{align}
There are flows from the $\mathcal{N}=4$ vacuum to either one of the $\mathcal{N}=2$ vacua with $\chi=0$ \eqref{eq:N2vacuum1} or $\chi=\pi/2$ \eqref{eq:N2vacuum1}.
These flows have a constant value for the scalar $\chi$ and can be constructed numerically in a way very similar to the one described at the end of Section~\ref{sec:oneN=2vacuum}. On the other hand, in order to flow from the vacuum in \eqref{eq:N2vacuum1} to the one in \eqref{eq:N2vacuum2} the scalar $\chi$ has to flow. This seems to imply that the numerical integration of the BPS flow equations is finely tuned and it is more challenging to construct these flows numerically. This is most likely related to the fact that both vacua in \eqref{eq:N2vacuum1} and \eqref{eq:N2vacuum2} are saddle points of the potential $V$.

%%%%%%%%%%%%%%%
\section{Discussion}
\label{sec:Discussion}
%%%%%%%%%%%%%%%

In this paper we studied the general structure of supersymmetric AdS vacua in half-maximal five-dimensional gauged supergravity as well as possible supersymmetric domain-wall solutions that connect them. Our results have a direct application to holography where they translate into constraints on the possible conformal vacua and RG flows of four-dimensional $\mathcal{N}=2$ SCFTs with a gravity dual.

The approach we took in this work is ``bottom-up'', i.e. we eschewed any reference to a particular embedding of the gauged supergravity into string or M-theory and studied the general structure of the five-dimensional theory. On one hand this allowed us to obtain very general results that should be applicable to all four-dimensional $\mathcal{N}=2$ SCFTs with a holographic dual, but on the other hand leaves the question open to what are concrete realizations in ten or eleven dimensions. For instance the domain wall connecting two supersymmetric AdS$_5$ vacua with sixteen supercharges studied in Section~\ref{sec:FlowN=4toN=4} should imply a corresponding RG flow connecting two $\mathcal{N}=2$ SCFTs with a gravity dual. We provided further evidence for this claim with the anomaly calculation in Section~\ref{sec:QFT}, however we are not aware of an explicit example of such an RG flow either in a ``top-down'' model arising from string or M-theory or in field theory. A potential realization of this $\mathcal{N}=2$ RG flow might be provided by the theories of class $\mathcal{S}$, i.e.\ $\mathcal{N}=2$ SCFTs arising from M5-branes compactified on a punctured Riemann surface, discussed in \cite{Gaiotto:2009gz}. The vev deformation of the UV $\mathcal{N}=2$ SCFTs which reduces the $\SU(2)_R$ R-symmetry and the $\SU(2)_F$ flavor symmetry to the diagonal subgroup (preserved all along the flow) may be provided by an appropriate ``Higgsing of a puncture'' on the Riemann surface. It was furthermore shown in \cite{Gaiotto:2009gz} how to describe this class of strongly interacting $\mathcal{N}=2$ SCFTs holographically in M-theory. What is missing to connect this set-up to our results is a well-defined prescription to assign a given five-dimensional gauged supergravity theory to any of the AdS$_5$ eleven-dimensional solutions in \cite{Gaiotto:2009gz}. It will be interesting to understand how to make such a link. We should also stress that the results presented in Section~\ref{sec:QFT} for the conformal anomalies of the UV and IR $\mathcal{N}=2$ SCFTs are valid beyond the supergravity approximation. It may be useful to emphasize that the IR central charges $a_{\rm IR}$ and $c_{\rm IR}$ in Section~\ref{sec:QFT} are those of the full IR SCFT. As a consequence of the partial spontaneous breaking of the UV global symmetry, the IR theory will contain a free sector made of Goldstone bosons in addition to the interacting sector.\footnote{We thank Prarit Agarwal for useful discussions on this.} In class $\mathcal{S}$ theories it is known how to separate the contributions of the Goldstone bosons from the rest, see e.g.~\cite{Tachikawa:2015bga}.

We were also able to describe general constraints for the existence of AdS$_5$ vacua and domain-walls with eight supercharges in a gauged supergravity theory with at least one AdS$_5$ vacuum with 16 supercharges. These results should be useful to understand RG flows between $\mathcal{N}=2$ and $\mathcal{N}=1$ SCFTs in four dimensions. The model with two vector multiplets discussed in Section~\ref{sec:oneN=2vacuum} is a particularly simple example of our general results which nevertheless is rich enough to capture interesting physics. For $\rho=2$ this model provides a holographic realization of the universal field theory RG flow discussed in \cite{Tachikawa:2009tt}. A well-known ``top-down'' example of this RG flow is provided by the $\mathcal{N}=1$ mass deformation of $\mathcal{N}=4$ SYM \cite{Girardello:1998pd,Freedman:1999gp}, as well as its $\mathbb{Z}_k$ orbifold \cite{Klebanov:1998hh,Corrado:2002wx,Corrado:2004bz}. It is widely expected that  this universal RG flow should connect also the $\mathcal{N}=2$ and $\mathcal{N}=1$ Maldacena-N\'u\~nez SCFTs arising from M5-branes wrapping a smooth Riemann surface \cite{Maldacena:2000mw}. These theories have holographic dual AdS$_5$ vacua but there is no known domain wall solution connecting them. The supergravity solution with $\rho=2$ described in Section~\ref{sec:oneN=2vacuum} should serve as a five-dimensional effective description of this holographic RG flow. It will certainly be very interesting to embed this five-dimensional model into a consistent truncation of eleven-dimensional supergravity. We are not aware of an explicit embedding of the model with $\rho\neq 2$ in Section~\ref{sec:oneN=2vacuum} into string or M-theory. However it is natural to conjecture that it may be describing holographic RG flows between the $\mathcal{N}=2$ Maldacena-N\'u\~nez SCFT and one of the $\mathcal{N}=1$ SCFTs with $0<|z|<1$ studied in \cite{Bah:2011vv,Bah:2012dg}. By the same token we can speculate that the model with one $\mathcal{N}=4$ and two $\mathcal{N}=2$ vacua described in Section~\ref{subsec:2N=2vacua} may describe holographic RG flows connecting the $\mathcal{N}=2$ Maldacena-N\'u\~nez vacuum with two of the $\mathcal{N}=1$ theories with $|z|<1$ in \cite{Bah:2011vv,Bah:2012dg}. To establish these conjectures rigorously one has to show how to construct a consistent truncation for the eleven-dimensional supergravity solutions of \cite{Bah:2011vv,Bah:2012dg} to five-dimensional gauged supergravity. Partial progress in this direction was presented in \cite{Szepietowski:2012tb}, however the solution to the full problem is still out of reach.

Finally we would like to point out that various interesting conjectures about the structure of RG flows in quantum field theory were presented in \cite{Gukov:2015qea,Gukov:2016tnp}. Supersymmetric CFTs with holographic duals and the RG flows connecting them provide a natural playing ground to explore these conjectures and we hope that some of our results may be useful in this context.

%%%%%%%%%%%%%
\section*{Acknowledgments}

We would like to thank Prarit Agarwal, Chris Beem, Sergio Benvenuti, Fri\dh rik Gautason and Parinya Karndumri for useful discussions. We are particularly grateful to Nick Halmagyi for participating at the initial stages of the development of this project and for many important discussions. The work of NB is supported in part by an Odysseus grant G0F9516N from the FWO, by the KU Lueven C1 grant ZKD1118 C16/16/005, and by the Belgian Federal Science Policy Office through the Inter-University Attraction Pole P7/37. The work of H.T.\ was supported by the EPSRC Programme Grant EP/K034456/1.

%%%%%%%%%%%%%%
\appendix

%%%%%%%
\section{Uniqueness of half-maximal AdS solutions in various dimensions}
\label{app:uniqueness4d}
%%%%%%%

Half-maximal gauged supergravity theories in different dimensions share a very similar structure. Their matter content and their couplings are completely fixed by the number of vector multiplets and the embedding tensor specifying the gauge group.
Therefore a natural question is the possible existence of a no-go result for multiple $\mathcal{N}=4$ vacua within half-maximal supergravity in dimension other than five, similar to the one obtained in Section~\ref{sec:uniquenessN=4_5d}. Indeed, in this appendix we show that, again under the assumption that the only compact subgroup of the gauge group is the R-symmetry of the vacuum, an analogous proof holds in dimensions four, six and seven. In more general situations it is natural to expect that there may be two distinct $\mathcal{N}=4$ vacua in four, six and seven dimensions. This should be viewed as a generalization of the five-dimensional results presented in Section~\ref{sec:twoN=2vacua}. It should then be possible to exhibit holographic RG flows connecting these distinct AdS vacua analogous to the ones studied in Section~\ref{sec:FlowN=4toN=4}. Indeed, examples of such flows in six- and seven-dimensional half-maximal gauged supergravity have been studied in \cite{Karndumri:2012vh,Karndumri:2014hma}. It will be interesting to study this further and understand these holographic RG flows from the point of the dual SCFT.

%%%%%%
\subsection{Four dimensions}
%%%%%%

In four dimensions, fully supersymmetric AdS vacua in half-maximal supergravity have been discussed in~\cite{Louis:2014gxa}.
There it was shown that the gauge group of $\mathcal{N}=4$ AdS vacua is
\begin{equation}
 G \,=\, H_+ \times H_- \times   H_{\rm c} \ \subset\ \SO(6,\mathfrak{n}) \ , \label{eq:4dN=4gaugegroup}
\end{equation}
so that $H_\pm$ have the same properties as $H_{\rm nc}$ in five dimensions, see \eqref{eq:N=4gaugegroup}, but with the novelty that $H_+$ and $H_-$ are electrically and magnetically gauged, respectively. In the AdS$_4$ vacuum we find again the breaking
\begin{equation}
H_\pm \to \SU(2)_\pm \ .
\end{equation}
In the holographically dual 3d $\mathcal{N}=4$ SCFT, $\SU(2)_{+}\times \SU(2)_{-}$ is the R-symmetry group. $H_{\rm c}$ is again compact and semi-simple and is gauged under vector multiplet gauge bosons. It corresponds to the group of flavor symmetries in the dual SCFT. The embedding tensor has components $f_\pm^{MNP}$ (while $\xi_\pm^M$ have to vanish in the $\mathcal{N}=4$ vacuum). If we define
\begin{equation}
\mathfrak{f}^{mnp}\ =\ \cV_M{}^m \cV_N{}^n \cV_P{}^p (\tau f_{-}{}^{MNP} - f_{+}{}^{ MNP}) \ ,
\end{equation}
where $\tau$ is the $\SL(2)$ complex scalar in the gravity multiplet,
then the $\mathcal{N}=4$ supersymmetry conditions read
\begin{align}
\cV_M{}^m \cV_N{}^n \cV_P{}^a f_{\pm}{}^{MNP} = 0 \ , \label{eq:4dN=4A}\\
\tfrac{1}{6}\varepsilon^{mnpqrs} \mathfrak{f}_{qrs} = - \iu \mathfrak{f}^{mnp} \ .
\end{align}
Using the quadratic constraints and the symmetries of the scalar manifold one can take
\be
\mathfrak{f}_{123}=\tfrac{1}{3\sqrt{2}} \mu \ ,\qquad
\mathfrak{f}_{456}= -\tfrac{1}{3\sqrt{2}} \iu  \mu \ .
\ee
The cosmological constant is $V=-\tfrac32\mathfrak{f}^{mnp} \mathfrak{f}^*_{mnp} = -|\mu|^2 $.

Let us fix one $\mathcal{N}=4$ AdS$_4$ vacuum to be at the origin, and let us assume that $H_c$ is trivial. Then we can argue analogously to the five-dimensional case that because of \eqref{eq:4dN=4A}, the following identities hold (up to $\SO(6)$ rotations) in the second vacuum
\begin{equation}\begin{array}{lll}
\cV_M{}^1\, = \, \Lambda_M{}^N  \delta_N{}^1  \ , &\qquad& \cV_M{}^4 \, = \, \tilde \Lambda_M{}^N  \delta_N{}^4  \ , \\
\cV_M{}^2\, = \, \Lambda_M{}^N  \delta_N{}^2  \ ,  && \cV_M{}^5\, = \,  \tilde \Lambda_M{}^N  \delta_N{}^5  \ , \\
\cV_M{}^3\, = \, \Lambda_M{}^N  \delta_N{}^3 \ ,&& \cV_M{}^6\, = \, \tilde \Lambda_M{}^N  \delta_N{}^6  \ ,
\end{array} \end{equation}
where $\Lambda$ and $\tilde \Lambda$ describe the embedding of ${\rm SU}(2)_\pm$ into $H_\pm$, respectively, which correspond to Goldstone directions in $\cV_M{}^m$, cf.\ \eqref{embedding_matrix}.
Note that the two SU(2) gauge groups cannot mix since they are electrically and magnetically gauged.

%%%%%%%%
\subsection{Six dimensions}
%%%%%%%%

In six dimensions, half-maximally supersymmetric AdS vacua are the only allowed supersymmetric AdS vacua and have been constructed and studied in \cite{Romans:1985tw,DAuria:2000afl,Andrianopoli:2001rs,Karndumri:2016ruc}. Let us start by briefly reviewing \cite{Karndumri:2016ruc}.

The gauge group is
\begin{equation}
 G \,=\, H \times H' \ \subset\ \SO(4,\mathfrak{n}) \ , \label{eq:6dN=4gaugegroup}
\end{equation}
where $H \subset \SO(3,\mathfrak{m})$ and $H' \subset \SO(1,\mathfrak{n}-\mathfrak{m})$ for some $\mathfrak{m}\leq\mathfrak{n}$. As in lower dimensions, this gauge group is spontaneously broken in a supersymmetric vacuum to its maximal compact subgroup, which turns out to be
\begin{equation}
\SO(3) \times H_c \ ,
\end{equation}
where $\SO(3)$ is gauged by three of the four graviphotons and corresponds to the R-symmetry group of the dual CFT, while $H_c \subset \SO(\mathfrak{n}-\mathfrak{m})$ corresponds to flavor symmetries.

The supersymmetry constraints on the embedding tensor reflect the discussion of the gauge group and are given by
\begin{align}
\cV_M{}^m \cV_N{}^n \cV_P{}^0 f{}^{MNP} &= 0 \ , \nn\\
\cV_M{}^m \cV_N{}^0 \cV_P{}^a f{}^{MNP} &= 0 \ , \nn\\
\cV_M{}^m \cV_N{}^n \cV_P{}^a f{}^{MNP} &= 0 \ , \nn\\
\cV_M{}^m \cV_N{}^n \cV_P{}^p f{}^{MNP} &= g\, \varepsilon^{mnp} \label{eq:6dN=4A}\ ,
\end{align}
for $m=1,2,3$. The gauge coupling $g$ and the mass $\tilde{m}$ of the two-form in the gravity multiplet together also determine the cosmological constant via
\begin{equation}
V = - 20 \,\tilde{m}^2 \left( \tfrac{g}{3\tilde{m}} \right)^{3/2} \ .
\end{equation}

Again, if we fix one half-maximal AdS$_6$ vacuum to sit at the origin and we assume that $H_c$ is empty, we can argue from the third equation in \eqref{eq:6dN=4A} that the vielbein $(\cV_M{}^0, \cV_M{}^m, \cV_M{}^a)$ of any other $\mathcal{N}=4$ AdS$_6$ vacuum must be related by the following embedding
\begin{align}
\cV_M{}^0\, = &\ \Sigma_M{}^N  \delta_N{}^0  \ , \nn\\
\cV_M{}^1\, = &\ \Lambda_M{}^N  \delta_N{}^1  \ , \nn\\
\cV_M{}^2\, = &\ \Lambda_M{}^N  \delta_N{}^2  \ , \nn\\
\cV_M{}^3\, = &\ \Lambda_M{}^N  \delta_N{}^3 \ ,
\end{align}
where $\Lambda$ has the same form as in \eqref{embedding_matrix} and therefore describes the embedding of $\SO(3)$ into $H'$. Similarly, $\Xi$ is a transformation in $\SO(1,\mathfrak{n}-\mathfrak{m})$ whose non-vanishing components are $\Xi_0{}^a$ and $\Xi_a{}^0$  given by
\begin{equation}
\Xi_0{}^a \,=\, {\tilde \lambda}^b f_{b0}{}^a  \ .
\end{equation}
Again, the transformations $\Lambda$ and $\Xi$ are precisely the Goldstone modes of the model, and thus the $\mathcal{N}=4$ vacuum is unique.

When $H_c$ contains an $\SO(3)$ subgroup, multiple supersymmetric AdS$_6$ solutions preserving all sixteen supercharges can be found. A supersymmetric flow between two such solutions was constructed in \cite{Karndumri:2012vh}.

%%%%%%%%
\subsection{Seven dimensions}
%%%%%%%%

Supersymmetric AdS vacua of half-maximal supergravity in seven dimensions have been discussed in \cite{Louis:2015mka}. Analogous to lower dimensions, the gauge group is of the form
\begin{equation}
   G = H \times H_c \subset \SO(3,\mathfrak{n}) \ ,
\end{equation}
where $H$ is spontaneously broken in the AdS vacuum  to its maximal compact subgroup $\SO(3)$, which is gauged by graviphotons and corresponds to the R-symmetry of the dual CFT. The compact group $H_c \subset \SO(\mathfrak{n})$ corresponds to flavor symmetries in the CFT.
This result is found by inspecting the supersymmetry conditions imposed on the embedding tensor components $f_{MNP},\xi_M$. These read:
\begin{align}
\xi_M &= 0\ ,\nn\\
\cV_M{}^m \cV_N{}^n \cV_P{}^a f{}^{MNP} &= 0 \ , \nn\\
\cV_M{}^m \cV_N{}^n \cV_P{}^p f{}^{MNP} &= g \,\varepsilon^{mnp} \ ,
\end{align}
where the gauge coupling constant $g$ determines the cosmological constant. Again, if $H_c$ is trivial the only transformations that leave these conditions invariant are
\begin{align}
\cV_M{}^1\, = &\ \Lambda_M{}^N  \delta_N{}^1  \ , \nn\\
\cV_M{}^2\, = &\ \Lambda_M{}^N  \delta_N{}^2  \ , \nn\\
\cV_M{}^3\, = &\ \Lambda_M{}^N  \delta_N{}^3 \ ,
\end{align}
with $\Lambda$ given by \eqref{embedding_matrix}, which corresponds to shifts by a Goldstone boson, establishing uniqueness of the supersymmetric AdS$_7$ vacuum.

Also in this case, when $H_c$ contains an $\SO(3)$ subgroup, one can have multiple AdS$_7$ solutions preserving sixteen supercharges, as well as supersymmetric flows connecting them, see \cite{Karndumri:2014hma} for an example.

%%%%%%%
\section{The generator of the IR $\U(1)_R$ symmetry}
\label{app:genTW}
%%%%%%%

In this appendix we show that the generator of the $\U(1)$ R-symmetry at the IR fixed point of the holographic flow discussed in Section~\ref{sec:oneN=2vacuum} is given in field theory units by
\be\label{eq:RIRapp}
R^{\rm IR}_{\mathcal{N}=1} = \frac{\rho}{\rho+2} R_{\mathcal{N}=2} + \frac{4}{\rho+2}I_3\ .
\ee

We can extract the information we need from the action of the supergravity gauge covariant derivative on the spinor parameter $\epsilon_i$.
The general form of the gauge covariant derivative was given in eq.~\eqref{s2:DMMN}.
When acting on the spinor parameter, this reads:
\be\label{CovDerSpinor}
D \epsilon \,=\, \nabla \epsilon - \tfrac{1}{4}(- \hat A^m \hat{f}^{mnp} \Gamma_{np} + \hat A^a \hat{f}^{anp} \Gamma_{np} + A^0 \hat\xi^{np}\Gamma_{np}) \epsilon\ ,
\ee
where $\nabla$ is the covariant derivative in the ungauged supergravity theory and we are suppressing the $\USp(4)$ indices on the spinor as well as on the $\SO(5)$ gamma matrices.

Before coming to the IR vacuum, let us consider the vacuum at the origin of the scalar manifold, preserving sixteen supercharges. Recalling the form \eqref{embtensorN=2vacuum}  of the embedding tensor, we have at that point:
\be
D \epsilon \,=\, \nabla \epsilon - \left(- \tfrac{1}{4}g A^m {\varepsilon}^{mnp} \Gamma_{np} -\tfrac{g}{2\sqrt2} A^0 \Gamma_{45}\right) \epsilon\ ,
\ee
where in this equation the indices $m,n,p$ run over $1,2,3$ only.
The embedding of the $\SU(2)\times\U(1)$ R-symmetry of the $\mathcal{N}=4$ vacuum in $\USp(4)$ is such that we have the following identification:
\be
\Gamma_{45} = R_{\mathcal{N}=2}\ ,\qquad -\tfrac{1}{4}\varepsilon_{mnp}\Gamma_{np} = I_m\ ,\quad m=1,2,3\ .
\ee
Therefore the covariant derivative can be written as
\be
D \epsilon \,=\, \nabla \epsilon - \left(g A^m I_m - \tfrac{g}{2\sqrt2} A^0 R_{\mathcal{N}=2}\right) \epsilon\ .
\ee

Now let us consider the supersymmetric flow discussed in Section \ref{sec:oneN=2vacuum}. Since we have found  there that $X_i{}^j = (\Gamma_3)_i{}^j $, the supersymmetries being preserved along the flow are
$\epsilon_+=\frac{1+\Gamma_3}{2}\epsilon$.
This also implies $\Gamma_{45}\epsilon_+ = -\Gamma_{12}\epsilon_+$.
Acting with the projector $\frac{1+\Gamma_3}{2}$ on \eqref{CovDerSpinor} to select these supersymmetries and using the expression for the dressed components of the embedding tensor given in \eqref{eq:flowingcouplings}, we arrive at
\be
 D \epsilon_+ =  \nabla \epsilon_+ - \tfrac{g}{2}\left(-  A^3 \cosh^2\phi + \tfrac{1}{\sqrt2 \rho}A^0 (\rho-2\sinh^2\phi) \right)\Gamma_{12} \epsilon_+  \ .
\ee
At the UV vacuum $\phi=0$ and this yields
\be\label{Depsilon+UV}
D \epsilon_+ \,=\, \nabla \epsilon_+ - (g A^3 I_3 - \tfrac{g}{2\sqrt2} A^0 R_{\mathcal{N}=2}) \epsilon_+\ .
\ee
Of the two symmetries generated by $R_{\mathcal{N}=2}$ and $I_3$, one linear combination is preserved along the flow, while another one is spontaneously broken, with the associated gauge field becoming massive. The symmetry that is preserved is manifest by evaluating the covariant derivative at the IR vacuum. Recalling that the latter is characterized by $\cosh^2\phi = \frac{\rho+2}{3}$, $\sinh^2\phi = \frac{\rho-1}{3}$,
we find
\be
 D \epsilon_+ =  \nabla \epsilon_+ - A^{\rm IR}\Gamma_{12} \epsilon_+ \ ,
\ee
with
\be
A^{\rm IR} = \tfrac{g}{6}(2+\rho)\left(\tfrac{1}{\sqrt2}\rho^{-1}A^0 - A^3\right)\ ,
\ee
and $\Gamma_{12}$ is the generator of the IR R-symmetry, which should be understood as the linear combination of $R_{\mathcal{N}=2}$ and $I_3$ we are after.  In addition, when in the main text we  discussed the gauge symmetries being broken, we found that the combination
\begin{equation}\label{eq:massive_vector}
  A^{\rm broken} = g(\sqrt{2} \rho^{-1} A^0 + A^3) \
\end{equation}
is massive (this is determined up to an overall normalization that will not matter).
Inverting the relation between $A^{\rm IR}, A^{\rm broken}$ and $A^{0},A^3$ we obtain
\begin{eqnarray}
  A^0 &=& \tfrac{\sqrt{2} \rho}{3g}  (A^{\rm broken} + \tfrac{6}{2+\rho} A^{\rm IR})  \ ,\nn \\
  A^3 &=& \tfrac{1}{3g} (A^{\rm broken} - \tfrac{12}{2+\rho} A^{\rm IR}) \ .
\end{eqnarray}
Plugging this in \eqref{Depsilon+UV}, we find that the generator multiplying $A^{\rm IR}$ is \eqref{eq:RIRapp}, which is what we wanted to show.

\bigskip

As an additional consistency check of our results, let us retrieve the ratio of central charges by studying the topological term in supergravity.
After ignoring all other vector fields, the relevant Chern-Simons term of half-maximal supergravity is
\begin{equation}
 {\cal L}_{\rm CS} \sim A^0 \wedge \diff A^3 \wedge \diff A^3 \ .
\end{equation}
If we also discard the vector becoming massive in the IR vacuum, the remaining Chern-Simons term is
\begin{equation}\label{eq:IR_CSrho}
{\cal L}^{\rm IR}_{\rm CS}  \sim \tfrac{32 \sqrt{2} \rho}{g^3(2+\rho)^3} A^{\rm IR} \wedge F^{\rm IR} \wedge F^{\rm IR}  \ .
\end{equation}
The coefficient of this term in the supergravity Lagrangian is proportional to the cubic R-symmetry anomaly of the IR superconformal R-symmetry which gives the leading contribution to the $a_{\rm IR}=c_{\rm IR}$ conformal anomaly.
The analogous Chern-Simons term in the UV can be obtained by setting $\rho\to1$ in \eqref{eq:IR_CSrho} to find
\begin{equation}\label{eq:UV_CSrho}
{\cal L}^{\rm UV}_{\rm CS}  \sim \tfrac{32 \sqrt{2} }{27g^3} A^{\rm UV} \wedge F^{\rm UV} \wedge F^{\rm UV}  \ .
\end{equation}
Taking the ratio of the two coefficients in \eqref{eq:IR_CSrho} and \eqref{eq:UV_CSrho} above we obtain the same result as the central charge ratio in \eqref{cratiosugra} computed by comparing the IR and UV values of the cosmological constant.

%%%%%%%%%%%%%%%%%%%%%%%%%%%%%%%%%%%%%%%%%%%%%%%%%%%%%%%%%%%%%%%%%%%%%%%%%%%%%%%%%%%%%%
\bibliographystyle{JHEP}
\bibliography{biblioN=4}

\end{document}